\documentclass[prb,showpacs,aps,twocolumn]{revtex4-1}

\usepackage{amssymb,amsmath}
\usepackage{graphicx}
\usepackage{multirow}

\usepackage[dvipsnames]{xcolor}
\usepackage{tikz,xstring}

\newcommand{\tarrow}[3]{\draw[#1,>=latex,NavyBlue,line width=20] (#2) -- (#3);}
\usetikzlibrary{backgrounds}
\newcommand{\sx}[3]{%
	\vcenter{\hbox{\scalebox{0.08}{
		\begin{tikzpicture}
			\usetikzlibrary{arrows}
			
			\def \Ox {4}
			\def \Oy {0}
			\def \Ix {8}
			\def \Iy {4}
			\def \IIx {0}
			\def \IIy {4}
			\def \amin {-2.5}		% bottom of arrow for spin up
			\def \amax {4}			% top ot arrow for spin up
			\def \dxspins {1.2}		% offset in x for spin downs
			\def \dyspins {-1.5}	% offset in y for spin downs
			
			\node (pos0) at (\Ox,\Oy) {};
			\node (pos1) at (\Ix,\Iy) {};
			\node (pos2) at (\IIx,\IIy) {};
			
			\node (0us) at (\Ox,\Oy+\amin) {};
			\node (0ue) at (\Ox,\Oy+\amax) {};
			\node (0ds) at (\Ox,\Oy+\amin+\dyspins) {};
			\node (0de) at (\Ox,\Oy+\amax+\dyspins) {};
			\node (02us) at (\Ox+\dxspins/2,\Oy+\amin) {};
			\node (02ue) at (\Ox+\dxspins/2,\Oy+\amax) {};
			\node (02ds) at (\Ox-\dxspins/2,\Oy+\amin+\dyspins) {};
			\node (02de) at (\Ox-\dxspins/2,\Oy+\amax+\dyspins) {};
			
			\node (1us) at (\Ix,\Iy+\amin) {};
			\node (1ue) at (\Ix,\Iy+\amax) {};
			\node (1ds) at (\Ix,\Iy+\amin+\dyspins) {};
			\node (1de) at (\Ix,\Iy+\amax+\dyspins) {};
			\node (12us) at (\Ix+\dxspins/2,\Iy+\amin) {};
			\node (12ue) at (\Ix+\dxspins/2,\Iy+\amax) {};
			\node (12ds) at (\Ix-\dxspins/2,\Iy+\amin+\dyspins) {};
			\node (12de) at (\Ix-\dxspins/2,\Iy+\amax+\dyspins) {};
			
			\node (2us) at (\IIx,\IIy+\amin) {};
			\node (2ue) at (\IIx,\IIy+\amax) {};
			\node (2ds) at (\IIx,\IIy+\amin+\dyspins) {};
			\node (2de) at (\IIx,\IIy+\amax+\dyspins) {};
			\node (22us) at (\IIx+\dxspins/2,\IIy+\amin) {};
			\node (22ue) at (\IIx+\dxspins/2,\IIy+\amax) {};
			\node (22ds) at (\IIx-\dxspins/2,\IIy+\amin+\dyspins) {};
			\node (22de) at (\IIx-\dxspins/2,\IIy+\amax+\dyspins) {};
			
			\begin{scope}
			\clip (\Ox-3.5,\Oy) rectangle (\Ox+3.5,\Oy+1.5);
			\draw[line width = 8] (pos0) ellipse (3 and 1);
			\end{scope}
			\begin{scope}
			\clip (\Ix-3.5,\Iy) rectangle (\Ix+3.5,\Iy+1.5);
			\draw[line width = 8] (pos1) ellipse (3 and 1);
			\end{scope}
			\begin{scope}
			\clip (\IIx-3.4,\IIy) rectangle (\IIx+3.5,\IIy+1.5);
			\draw[line width = 8] (pos2) ellipse (3 and 1);
			\end{scope}	
			
			\IfEqCase{#1}{%
				{0}{}
				{1}{\tarrow{->}{0us}{0ue}}
				{-1}{\tarrow{<-}{0ds}{0de}}
				{2}{\tarrow{<-}{02ds}{02de}\tarrow{->}{02us}{02ue}}
			}[\PackageError{sx}{Undefined option to dot: #1}{}]%
			\IfEqCase{#2}{%
				{0}{}
				{1}{\tarrow{->}{1us}{1ue}}
				{-1}{\tarrow{<-}{1ds}{1de}}
				{2}{\tarrow{<-}{12ds}{12de}\tarrow{->}{12us}{12ue}}
			}[\PackageError{sx}{Undefined option to dot: #2}{}]%
			\IfEqCase{#3}{%
				{0}{}
				{1}{\tarrow{->}{2us}{2ue}}
				{-1}{\tarrow{<-}{2ds}{2de}}
				{2}{\tarrow{<-}{22ds}{22de}\tarrow{->}{22us}{22ue}}
			}[\PackageError{sx}{Undefined option to dot: #3}{}]%
			
			\begin{scope}
				\clip (\Ox-3.5,\Oy-1.5) rectangle (\Ox+3.5,\Oy);
				\draw[line width = 8] (pos0) ellipse (3 and 1);
			\end{scope}
			\begin{scope}
				\clip (\Ix-3.5,\Iy-1.5) rectangle (\Ix+3.5,\Iy);
				\draw[line width = 8] (pos1) ellipse (3 and 1);
			\end{scope}
			\begin{scope}
				\clip (\IIx-3.4,\IIy-1.5) rectangle (\IIx+3.5,\IIy);
				\draw[line width = 8] (pos2) ellipse (3 and 1);
			\end{scope}
		\end{tikzpicture}
	}}}
}

\begin{document}

\title{Fano stability diagram of a symmetric triple quantum dot}

\author{Michael Niklas}
\affiliation{Institute for Theoretical Physics, University of
Regensburg, 93040 Regensburg, Germany}

\author{Andreas Trottmann}
\affiliation{Institute for Theoretical Physics, University of
Regensburg, 93040 Regensburg, Germany}

\author{Andrea Donarini}
\affiliation{Institute for Theoretical Physics, University of
Regensburg, 93040 Regensburg, Germany}
\author{Milena Grifoni}
\altaffiliation{E-mail: milena.grifoni@ur.de}
\affiliation{Institute for Theoretical Physics, University of
Regensburg, 93040 Regensburg, Germany}

\date{\today}


\begin{abstract}
The Fano factor stability diagram of a C$_{3v}$ symmetric triangular quantum dot is analysed for increasing electron fillings $N$. At low filling, conventional Poissonian and sub-Poissonian behavior is found. At larger filling, $N\ge 2$, a breaking of the electron-hole symmetry is manifested in super-Poissonian noise with a peculiar bias voltage dependence of the Fano factor at Coulomb and interference blockade. An analysis of the Fano map unravels a nontrivial electron bunching mechanism arising from the presence of degenerate many-body states combined with orbital interference and Coulomb interactions. An expression for the associated dark states is provided for generic $N$. 
\end{abstract}

\maketitle

\section{Introduction}
\begin{figure}[t!]
	\centering
	\includegraphics[width=\columnwidth]{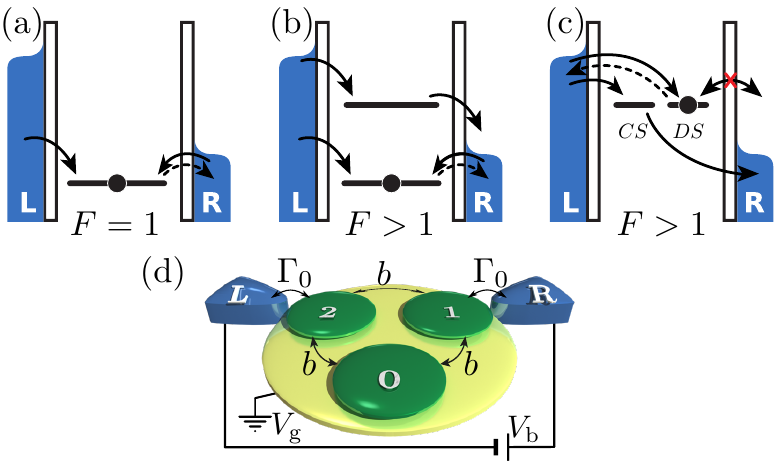}
	\caption{
	(a) Single level quantum dot in the Coulomb blockade regime with Fano factor $F=1$. (b) If an excited state is present in the bias window of a Coulomb blockaded quantum dot, electron bunching through the excited state yields super-Possonian noise $(F>1)$. (c) Interference of energy degenerate orbitals gives rise to coupled $(CS)$, and dark $(DS)$ states, and in turn to super-Poissonian noise. Solid arrows show fast processes, dashed ones the dominant slow processes. (d) A gated triangular triple quantum dot (TQD) features all the three dynamical situation sketched in (a)-(c).
	}
	\label{fig1}
\end{figure}

Current fluctuations in out of equilibrium nanoscale systems can yield information about relevant transport mechanisms not accessible from the knowledge of the average current only \cite{Landauer1998}. In fermionic tunneling structures the interplay between Pauli principle and repulsive Coulomb interactions usually yields Poissonian and sub-Poissonian noise, corresponding to a Fano factor $F=1$, and $ F<1$, respectively.
For example, in single level quantum dot systems one finds Poissonian shot noise at Coulomb blockade,  see Fig.~\ref{fig1}(a)), and sub-Poissonian noise with $1/2 < F<1$ in the sequential transport regime \cite{Hershfield1993, Hanke1993, Korotkov1994, Nauen2002, Bagrets2003, Thielmann2003}. The latter is an indication that each tunneling barrier can be regarded as an independent source of Poissonian noise \cite{Sukhorukov2001}.
The enhancement of the shot noise, i.e. $F>1$,  requires a multilevel structure of the quantum dot \cite{Sukhorukov2001, Cottet2004, Belzig2005} or complex multiple quantum dot devices \cite{Kießlich2003, Groth2006, Poltl2009, Schaller2009, Benito2016}.
Independent of the details of the nanosystems, super-Poissonian noise implies the presence of slow and fast channels, and mechanisms which occasionally allow for charge transfer on a time scale much shorter than the average residence time in the slow channel state, see e.g. Figs. 1(b) and 1(c). Thus super-Poissonian noise is a signature of fermionic bunching and in turn of subtle quantum correlations, being the topic of this work.

In this manuscript we investigate the Fano stability diagram of a C$_{3v}$ symmetric triangular triple quantum dot (TQD), schematically sketched in Fig.~\ref{fig1}(d), as a function of its occupation.
TQDs are the smallest systems where the interplay of statistics, Coulomb interactions and  geometry allows one the study of peculiar many-body effects  such as super-exchange induced triplet-singlet transition \cite{Korkusinski2007}, many-body interference \cite{Poltl2009, Kostyrko2009, Donarini2009}, cellular automata phenomena \cite{Pan2012}, charge frustration  \cite{Seo2013, Andergassen2013, Lee2013}, or channel blockade \cite{Kotzian2015}.
TQDs have been recently realized in lateral semiconducting heterostrucures \cite{Gaudreau2006, Pan2012, Seo2013, Kotzian2015}, which are tunable down to the few electron regime by means of plunger and depletion gates \cite{Gaudreau2006}, and by means of atomic STM manipulation \cite{Weitering2014}.
In the latter experiment, orbital degeneracy in a C$_{3v}$ symmetric triangular dot could be demonstrated.

So far, investigations of transport noise have been restricted to set-ups in which the C$_{3v}$ symmetry of an isolated TQD is broken in various ways, e.g. by assuming unequal interdot hoppings and/or onsite energies \cite{Groth2006, Poltl2009, Dominguez2010, Weymann2011, Wrzesniewski2015}. These asymmetries remove orbital degeneracies and hence the possibility to observe current suppression due to the destructive orbital interference \cite{Donarini2010}. We show that the Fano map of a C$_{3v}$ symmetric, weakly coupled TQD is characterized by regions of super-Poissonian noise \emph{outside} the Coulomb blockade region, see Fig.~\ref{fig1}(c), with specific fractional values of the Fano factor being a signature of such many-body interference.
By exploting an analytical solution for the TQD spectrum, the explicit form of the dark states is provided, including the most complex situation of half-filling.

\section{Model and spectrum}
We examine a single-electron transistor model described by the total Hamiltonian $H=H_\mathrm{TQD}+H_\mathrm{tun} + H_\mathrm{res}$. The TQD spectrum derives from a three-site Hamiltonian with hopping $b<0$, onsite Coulomb repulsion $U$ and inter-site repulsion $V$, 
\begin{align}
	\label{eq:hamiltonian}
	&H_\mathrm{TQD}
	=
	\xi \sum_{i\sigma} n_{i\sigma}
	+
	b  \sum_{i\not=j,\sigma} d_{j\sigma}^\dagger d_{i\sigma} +
	 \\
	&
	U  \sum_i \left( n_{i\uparrow} - \frac{1}{2} \right) \left( n_{i\downarrow} - \frac{1}{2} \right) 
	+V  \sum_{i<j} \left( n_i -1 \right) \left( n_j-1 \right)
	, \nonumber
\end{align}
where $\xi=\varepsilon -e \eta V_{\mathrm{g}}$ includes the onsite energy $\varepsilon$ and the effects of an applied gate voltage $V_\mathrm{g}$ with level arm $\eta$. Here, operators $d_{i\sigma}^\dagger$ and $d_{i\sigma}$ create and annihilate an electron with spin projection $\sigma$ in dot $i=0,1,2$, and $n_i=\sum_\sigma n_{i\sigma}$, $n_{i\sigma}=d_{i\sigma}^\dagger d_{i\sigma}$, cf. Fig.~\ref{fig1}(d).
The two leads are considered as reservoirs  at chemical potentials $\mu_{L/R}=\mu_0 \pm eV_\mathrm{b}/2$ for the left ($L$) and right ($R$) lead, with $V_\mathrm{b}$ the applied bias voltage, where we measure the energy from the equilibrium chemical potential $\mu_0=0$.
The corresponding Hamiltonian is
$H_\mathrm{res}=\sum_{\alpha k \sigma} \xi_{\alpha k} c_{\alpha k\sigma}^\dagger c_{\alpha k\sigma}$, with $\alpha=L,{R}$. Finally, tunneling between TQD and leads is described by $H_\mathrm{tun}=\sum_{\alpha k\sigma}\sum_i\left( t^*_{\alpha i}c_{\alpha k\sigma}^\dagger d_{i\sigma} +t_{\alpha i} d_{i\sigma}^\dagger c_{\alpha k\sigma} \right)$.
The single orbital approximation yields a realistic description of lateral TQD devices \cite{Pan2012, Seo2013, Gaudreau2006, Kotzian2015} as long as $eV_\mathrm{b}$ and $eV_\mathrm{g}$ are of the order of the hopping parameter $b$ \cite{Hsieh2012}. 
In the following, we consider equal coupling to the left and right leads, and set $t_{L2}=t_{R1}=t$ and otherwise $t_{\alpha i}=0$.
We identify for later convenience $d_{1\sigma}=d_{R\sigma}$, $d_{2\sigma}=d_{L\sigma}$.

The single particle part of the TQD Hamiltonian Eq.~(\ref{eq:hamiltonian}) is diagonalized in the basis of the angular momentum states  $\{\vert l \rangle =1/\sqrt{3} \sum_{j=0}^2 e^{-i j l 2\pi/3 } \vert j \rangle$, $\{l=0,\pm1\}$. Accounting for the spin degree of freedom $\sigma$, in the following we use this single particle basis to construct many-body states in the occupation number representation, where a generic vector $\vert n_{0\uparrow}, n_{1\uparrow}, n_{-1\uparrow}; n_{0\downarrow}, n_{1\downarrow}, n_{-1\downarrow} \rangle $ is fully characterized by the occupation numbers $n_{l\sigma}$. Finally, we use this many-body basis to diagonalize the TQD Hamiltonian and find its eigenvalues and eigenfunctions. Several symmetries have been exploited in the analytical diagonalization: $H_\mathrm{TQD}$ commutes in fact with the total particle number operator $N=\sum_{l\sigma}n_{l\sigma}$, the total spin operator $S^2=\sum_{i\,l\sigma\sigma^\prime}(d^\dagger_{l\sigma}s^i_{\sigma\sigma^\prime}d_{l\sigma^\prime})^2$ (here is $s^i=\frac{\hbar}{2}\sigma^i$ and $\sigma^i$ the $i$-th Pauli matrix), the spin projection $S_z=\frac{\hbar}{2}\sum_{l\sigma}\sigma n_{l\sigma}$, and the angular momentum operator \cite{Kostyrko2009} $L_z=\hbar\sum_{l\sigma} l n_{l\sigma}\vert_{\mathrm{mod}\; 3}$. By ordering the many-body states according to the quantum numbers $N$, $S$, $S_z$ and $L_z$ associated to these operators, we could reduce the Hamiltonian into a blockdiagonal form with blocks of maximal size $3\times3$, and then complete the diagonalization. The set of eigenvalues listed above together with the energy $E$ fully characterize the eigenvectors of the interacting TQD Hamiltonian, a crucial knowledge for the forthcoming analysis.
In the following we use the notation $\vert N, E; S, S_z, L_z \rangle$ or $\vert N,\alpha_i, L_z \rangle$, with $\alpha_i=\{E_{N_i}; S, S_z\}$, for a generic eigenvector.
In particular, $S^2\vert N, E; S, S_z, L_z \rangle=\hbar^2 S(S+1)\vert N, E; S, S_z, L_z \rangle$ and, as usual, $-S\le S_z \le S $.
The C$_{3v}$ group of the TQD comprises also three reflection planes perpendicular to the system. In particular, we introduce the reflection operator $\sigma_{v0}$ such that $\sigma_{v0}d_{1\sigma}^\dagger \sigma_{v0}=d_{2\sigma}^\dagger$ and $\sigma_{v0}d_{0\sigma}^\dagger \sigma_{v0}=d_{0\sigma}^\dagger$. The overall phase of the eigenstates is taken in such a way that $\sigma_{v0}\vert N, \alpha_i, 1\rangle = \vert N, \alpha_i, -1\rangle$.

Such eigenvectors and the associated eigenvalues in the occupation number basis are reported in Appendix \ref{sec:A-spectrum}.
For convenience we set $\hbar=1$ in the quantum numbers.
\begin{figure}[t]
	\includegraphics[width=\columnwidth]{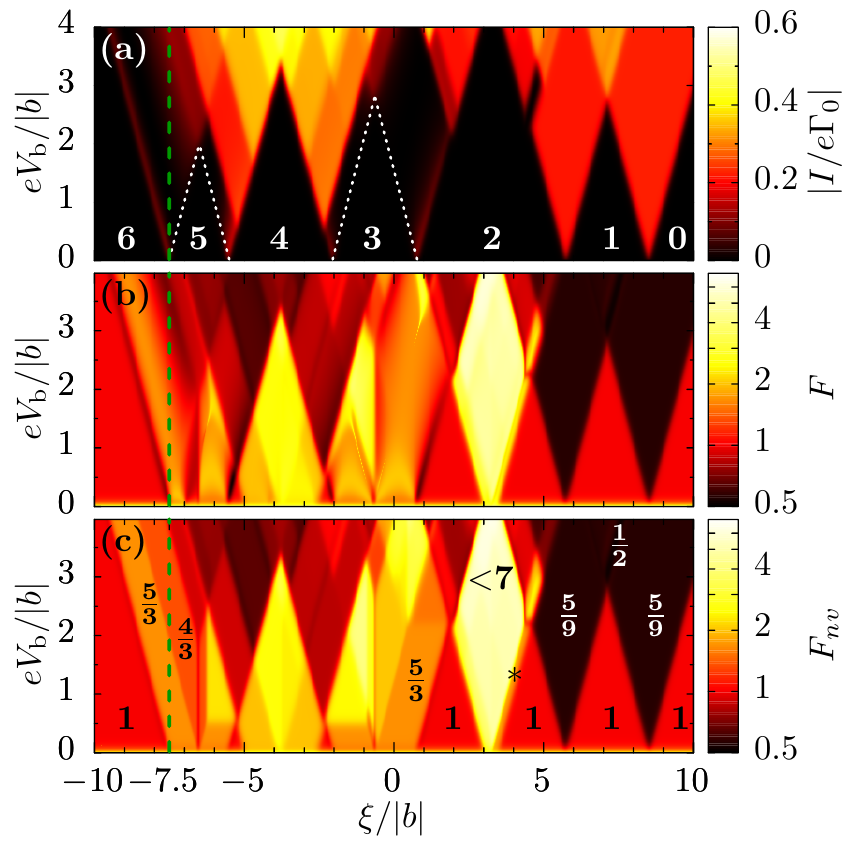}
	\caption{Current and Fano factors vs appplied gate and bias voltages. (a) Average current where the number of electrons in the blockade regions is displayed. The white dotted lines delimit regions where transport is inhibited due to Coulomb blockade. (b) Fano factor and (c) Fano factor without Lamb shifts due to virtual transitions. Some values discussed in the text are indicated. Parameters used for the simulations are
$U=5\vert b\vert$, $V=2\vert b\vert$, $k_\mathrm{B}T=0.002 \vert b\vert$, $k_\mathrm{B}T=20\Gamma$ and $b<0$.
	}
	\label{fig2}
\end{figure}

\section{Current and noise in a reduced density matrix approach}
To compute the current and shot noise we use a master equation approach for the generalized reduced density matrix $\rho_\chi=\mathrm{Tr}_\mathrm{res}\left\{ e^{i\chi N_\mathrm{R}}\rho\right\}$, where $\chi$ and $N_\mathrm{R}$ are the counting field and number operator for the right lead, and $\rho$ the total density operator \cite{Bagrets2003, Flindt2010}. A truncation to second order in $H_\mathrm{tun}$ yields the generalized master equation $	
	\dot{\rho}_\chi
	=
	\left[
		\mathcal{L}
		+(e^{i\chi}-1)\mathcal{J}^+
		+(e^{-i\chi}-1)\mathcal{J}^-
	\right]\rho_\chi
$ \cite{Kaiser2007}, where $\mathcal{L}$ is the Liouville superoperator, and we defined the current superoperators for increasing, $\mathcal{J}^+$, and decreasing, $\mathcal{J}^-$, the number of electrons in the right lead.
This results in the equations for the stationary reduced density matrix, $\rho^\infty=\lim_{t\to\infty}\rho_{\chi=0}$, and the moments $\mathcal{F}_k^\infty=\lim_{t\to\infty} d^k/d(i\chi)^k\rho_\chi\vert_{\chi=0}$. Introducing the traceless part of the first moment, $\mathcal{F}_{1\perp}^\infty = (1-\rho^\infty\mathrm{Tr}_\mathrm{TQD})\mathcal{F}_1^\infty$, one finds in particular
\begin{equation}
	\begin{split}
		\mathcal{L}\rho^\infty
		&	=
		-\frac{i}{\hbar}\left[H_\mathrm{TQD}+H_\mathrm{LS},\rho^\infty \right]+\mathcal{L}_t\rho^\infty= 0,
		\\
		\mathcal{L}\mathcal{F}_{1\perp}^\infty
		&	=
		\left(-I/e-\mathcal{J}^++\mathcal{J}^-\right)\rho^\infty,
	\end{split}
	\label{eq:liouvillian}
\end{equation}
where $\mathcal{L}_t$ is the tunneling part of the Liouvillian.
The Lamb shift Hamiltonian \cite{Schultz2009, Donarini2009} $H_\mathrm{LS}$ generates a precession dynamics within orbitally degenerate subspaces. Its excplicit form is given in Sec.~\ref{sec:2-3}. The operatorial form of Eq.~(\ref{eq:liouvillian}) fully accounts for interference effects captured in the off-diagonal elements of $\rho^{\infty}$.
The current $I$ (first cumulant) and shot noise $S$ (second cumulant) in turn follow as \cite{Kaiser2007}
\begin{align}
	I
	&	=
	-e~\mathrm{Tr}_\mathrm{TQD}\left(\mathcal{J}^+-\mathcal{J}^-\right)\rho^\infty,
	\\
	S
	&	=
	e^2\mathrm{Tr}_\mathrm{TQD}\left[
		2\left(\mathcal{J}^+-\mathcal{J}^-\right)\mathcal{F}_{1\perp}^\infty
		+\left(\mathcal{J}^++\mathcal{J}^-\right)\rho^\infty
	\right]
	.\nonumber
	\label{eq:IS}
\end{align}
As a dimensionless measure for the relative noise strength we employ the Fano factor $F=S/e|I|$.

\section{Current and Fano maps}
The stationary current is shown as a function of bias and gate voltage in Fig. 2(a). For comparison, the same parameters as in the work by A.~Donarini et al. \cite{Donarini2010} were used.
Notice that the closed geometry of the TQD breaks the particle-hole symmetry otherwise present in linear triple dots \cite{Korkusinski2007}.
The stability diagram displays Coulomb diamonds inside which current is exponentially suppressed (in second order in $H_\mathrm{tun}$) due to Coulomb blockade, but also regions outside the Coulomb diamonds with suppression due to orbital interference \cite{Donarini2009}. Coulomb diamonds are indicated with dotted lines when no longer visible due to the additional interference blockade.
A measurement of the current alone, however, does not enable one to tackle down the different blockade mechanisms. In contrast, the Fano map, shown  in Fig.~\ref{fig2}(b), displays a much richer structure than the current. In Fig.~\ref{fig2}(c) we  show the Fano factor $F_{nv}$, which is obtained by neglecting the Lamb shift term $H_\mathrm{LS}$ in  Eq. (\ref{eq:liouvillian}). Clearly, the virtual transitions responsible for the Lamb shifts \emph{blur} the otherwise poligonal Fano pattern.

At first glance one can observe a sub-Poissonian shot noise $1/2<F<1$ in the transport regime and both Poissonian, $F=1$, and super-Poissonian, $F>1$, shot noise in the regions of vanishing current. Furthermore, $F$ diverges when $V_\mathrm{b}\to 0$ due to Johnson-Nyquist noise.
Finally, vertical steps in the Fano factor are clearly visible at the center of the $3$- and $5$-particles Coulomb diamonds.
At these positions the energy levels of the states with one electron more and less than the participating Coulomb- or interference-blocked state are aligned, and a little change in the gate voltage favors one or the other side in transport, leading to a sudden change in the statistics that is unaffected by the bias voltage \cite{Bodoky2008, Belzig2014}.

The complexity of the Fano pattern increases with growing electron filling, so from right to left. The breaking of electron-hole symmetry is strikingly revealed in a Fano factor smaller (larger) than one,  in the transport (blockade) regions involving the $N=0 \,(6)$ and $N=1\, (5)$ groundstates.
Moreover, large values of $F$ are observed for intermediate filling.
The Fano map at low filling is easily understood by observing that the one-electron groundstates $\{|1, E_{1_0}; 1/2, \pm 1/2, 0 \rangle \}$ are only spin degenerate. Then, in the region with zero and one electron occupation, the Fano map resembles the one of the single impurity Anderson model, with $F=5/9$ and $F=1/2$ in the transport regions, and $F=1$ at Coulomb blockade \cite{Thielmann2003}.
At larger filling $N\ge 2$,  super-Poissonian noise signals  the presence of fast $(f)$ and slow $(s)$ channels. In this situation the Fano factor can be described in terms of effective filling rates $\Gamma_\alpha^p= R^p_\alpha\Gamma_{0\alpha} $, $p=s,f$, as shown in Appendix \ref{sec:A-Fnv}, as
\begin{equation}
	F_{nv}=1+\frac{2\Gamma^f_{L}}{\Gamma^s_{L}+\Gamma^s_{R}}, \quad \mu_L>\mu_R,
	\label{eq:fanob}
\end{equation}
where $\Gamma_{0\alpha}=2\pi \vert t\vert^2D_\alpha/\hbar$ is the bare tunneling rate for lead $\alpha$, proportional to the density of states at the Fermi energy $D_\alpha$.
In the following we assume identical leads, such that $\Gamma_{0L}=\Gamma_{0R}=\Gamma_0$, see Fig. \ref{fig1}(d). The coefficients $R^p_\alpha$ weight the fast and slow channels, and account for both spin degeneracies and orbital interference. The latter is ubiquitous in our symmetric TQD. Hence, even at Coulomb blockade, the observed values of $F$ and $F_{nv}$ cannot be simply explained in terms of the  channel blockade mechanism \cite{Belzig2005}. This requires a Coulomb blockaded level, and excited states in the transport window which provide a fast transport channel, see  Fig.~\ref{fig1}(b).
The larger the excited states degeneracy, the larger is the Fano factor.
Consider the Coulomb diamond with $N=2$ at the bias value indicated by the symbol * in Fig.~\ref{fig1}(c), where only the  groundstate $\vert 2_0\rangle\equiv \vert 2, E_{2_0}; 0,0,0 \rangle$, and the first set of excited levels  given by the sextuplet $\{ \vert 2, E_{2_1}; 1, S_z, \pm1 \rangle, S_z=0,\pm 1\}$  enter the transport window.
By applying Eq.~(\ref{eq:fanob}) naively assuming that $R^p_\alpha$ is just the channel multiplicity ($ R^s_\alpha=1$ and $R^f_\alpha =6$), one predicts $F_{nv}=7$, which is bigger than the observed value $F_{nv}\approx 2$.
At higher bias, transitions to the states $\vert 2, E_{2_2}; 0,0,\pm1 \rangle$ and $\vert 3, E_{3_0}; 1/2,\pm1/2,\pm1 \rangle$ enter the bias window and even larger values of $F_{nv}$ are expected. This is not observed in our TQD, where $F_{nv}< 7$. However, the sheer amount of open transitions makes analytics practically impossible, especially since three of these states are orbitally degenerate.

At interference blockade, with the blocking state being decoupled at the right lead, cf. Fig.~\ref{fig1}(c), Eq.~(\ref{eq:fanob}) holds with $\Gamma^s_R=0$.
For transitions to the left lead through a coupled and decoupled state one would naively expect $ R^f_L=R^s_L$ and therefore a Fano factor $F_{nv}=3$, as found in many systems \cite{Kießlich2003, Groth2006, Belzig2005, Cottet2004}. The value $F_{nv}=5/3$  observed e.g. at the resonance involving the groundstates $\vert 2_0 \rangle$ and $\{|3, E_{3_0};1/2,\pm 1/2,\pm 1 \rangle\}$, see Fig.~\ref{fig2}(c), again indicates that the evaluation of the weights $R^p_\alpha$ requires a precise analysis of interference with its associated ``dark states''.
In this respect we dedicate the next section to study the generic form of a dark state and give specific examples.

\section{Dark states of a C$_{3v}$ symmetric TQD}
When a set of orbitally degenerate levels participate in transport, interference can inhibit the escape from one many-body state with $N$ particles at one lead, such that electrons can leave this so called ``dark state'' (DS) only via thermal activation through the other lead or via virtual excitations, see Fig.~\ref{fig1}(c).
This yields current suppression. In the following we focus exemplarily on transitions blocked at the right lead and which involve an orbitally degenerate multiplet with $N$ particles and a singlet with $N-1$ particles. We denote the latter state, which necessarily has $L_z=0$, by $\vert N-1; 0 \rangle$, and define a DS through the requirement $\langle N-1; 0 | d_{1\sigma} |N; DS \rangle=0$.
Because a $L_z=0$ state and $d_{1\sigma}$ are invariant upon a reflection $\sigma_{v1}$ which leaves the site $1$ invariant and sends $2 \leftrightarrow 0$, such a blocking state must be antisymmetric under $\sigma_{v1}$. Expressing $d_{1\sigma}$ in the angular momentum basis, we find for the DS the anti-bonding linear combination
\begin{equation}
	\vert N, \alpha_i; DS \rangle
	=\!
	\frac{1}{\sqrt 2}\big[
		e^{i\frac{2\pi}{3}}\vert N, \alpha_i,1 \rangle - e^{-i\frac{2\pi}{3}}\vert N, \alpha_i,-1 \rangle 
		\big]
	\!
	, 
\end{equation}
where $\alpha_i =\{E_{N_i};S,S_z\}$  accounts for the energy and spin of the DS. Thus, a DS is an antibonding combination of states with angular momentum $L_z=\pm1$. Note that this result is independent of spin degrees of freedom. Indeed this state fulfills
\begin{align}
\label{eq:DS_proof}
	&\langle N-1; 0 \vert d_{1\sigma} \vert N; DS \rangle 
	=
	\nonumber \\ \nonumber
	&\langle N-1; 0 \vert \sum_l e^{-il2\pi/3} d_{l\sigma} \left[e^{i2\pi/3}\vert N; 1 \rangle - e^{-i2\pi/3}\vert N; -1 \rangle\right]
	\\ \nonumber
	&=
	\langle N-1; 0 \vert d_{l=1\sigma} \vert N; 1 \rangle - \langle N-1; 0 \vert d_{l=-1\sigma} \vert N; -1 \rangle
	\\ 
%	=	\langle N-1, \alpha_i, 0 \vert d_{l=1\sigma} \vert N, \alpha_j, 1 \rangle - \underbrace{\langle N-1, \alpha_i, 0 \vert \sigma_{v0}}_{\langle N-1, \alpha_i, 0 \vert}\underbrace{\sigma_{v0}d_{l=-1\sigma}\sigma_{v0}}_{d_{l=1\sigma}}\underbrace{\sigma_{v0} \vert N, \alpha_j, -1 \rangle}_{\vert N, \alpha_j, 1 \rangle}
%	\\ 
	&=
	0
	,
\end{align}
which shows that a transition is forbidden at the right lead.
To the bonding linear combination it corresponds the coupled state $\vert N; CS \rangle$.
Expressing the dark states in position basis $\{{0\uparrow}, {1\uparrow}, {2\uparrow}; {0\downarrow}, {1\downarrow}, {2\downarrow}\}$ yields further insight into the blocking mechanism.
Let us consider their composition for increasing electron filling.
The dark state for the one--particle first excited state with $S_z=1/2$, is
\begin{equation}
\label{eq:DS_1_1}
	\vert 1, E_{1_1}; \frac{1}{2}, \frac{1}{2}; DS \rangle
	=
	\frac{1}{\sqrt{2}}\left(\sx{1}{0}{0}-\sx{0}{0}{1}\right)
	,
\end{equation}
and similarly for the dark state with $S_z=-1/2$.
Thus we recover the familiar result by C.-Y.~Hsieh et al. \cite{Hsieh2012}, where the DS is a state without occupation of the right-coupled dot $1$.
On the other hand the vanishing of the matrix element in Eq.~(\ref{eq:DS_proof}) also comes naturally from the fact that the DS (\ref{eq:DS_1_1}) is \emph{antisymmetric} under the operation $\sigma_{v1}$ while both $d_{1\sigma}$ and the vacuum state $\vert 0, \alpha_0, 0 \rangle$ are symmetric.

For the two--particle first excited state with $S_z=1$ we obtain
\begin{equation}
\label{eq:DS_2_1_1}
	\vert 2, \alpha_{1};  DS \rangle
	=
	\frac{1}{\sqrt{6}}\left(\sx{0}{1}{1}+\sx{1}{1}{0}+2\sx{1}{0}{1}\right)
	,	
\end{equation}
and similarly for $S_z=-1$.
For the case $S_z=0$ we find
\begin{align}
\label{eq:DS_2_1_0}
	&\vert 2, E_{2_1}; 1, 0; DS \rangle
	=
	\frac{1}{2\sqrt{3}}\left[
		\left(\sx{0}{1}{-1}-\sx{-1}{1}{0}\right)
		\right. 
	\\ \nonumber 
		&\left. 
		- \left(\sx{0}{-1}{1}-\sx{1}{-1}{0}\right)
		+ 2\left(\sx{1}{0}{-1}-\sx{-1}{0}{1}\right)
	\right]
	.
\end{align}
The composition of states shown in Eqs.~(\ref{eq:DS_2_1_1}) and (\ref{eq:DS_2_1_0}) is counterintuitive because they admit \emph{finite} occupation of the dot $1$. However, again the vanishing of the transition amplitude (\ref{eq:DS_proof}) results from the fact that the DS is antisymmetric with respect to the reflection $\sigma_{v1}$ and the state $\vert 1, \alpha_0, 0 \rangle$ is symmetric. Notice that crucially the two contributions with single occupation of dot $1$ give a contribution of opposite sign to the amplitude (\ref{eq:DS_proof}).

The three--particles groundstate with $S_z=1/2$ is given by the intricated superposition
\begin{widetext}
\begin{align}
	\vert 3, E_{3_0}; \frac{1}{2}, \frac{1}{2}; DS \rangle
	=
	\frac{1}{3\sqrt{2}}\Bigg[
	&\left(v_{0,1}-2v_{0,0}+v_{0,-1}\right)
	\left(\sx{2}{1}{0}+\sx{0}{1}{2}\right)
	+
	\left(2v_{0,1}-v_{0,0}-v_{0,-1}\right)
	\left(\sx{2}{0}{1}+\sx{1}{0}{2}\right)
	\nonumber\\
	+&
	\left(v_{0,1}+v_{0,0}+v_{0,-1}\right)
	\left(\sx{-1}{1}{1}+\sx{1}{1}{-1}\right)
	+
	\left(v_{0,1}+v_{0,0}-2v_{0,-1}\right)
	\left(\sx{1}{2}{0}+\sx{0}{2}{1}\right)
	\nonumber\\
	+&
	2\left(v_{0,1}+v_{0,0}+v_{0,-1}\right)
	\sx{1}{-1}{1}
	\Bigg]
	,
\end{align}
\end{widetext}
where $v_{x,y}$ is given in the caption of Tab.~\ref{tab1}. Analogously the state with $S_z=-1/2$ can be constructed. Again the vanishing of the transition amplitude (\ref{eq:DS_proof}) results from a nontrivial quantum cancellation.

\section{Interference blockade at the $2_0\leftrightarrow 3_0$ resonance}
\label{sec:2-3}
We apply the results above to investigate the bias region involving transitions among an $N$-particles groundstate singlet, and an orbitally degenerate $(N+1)$-particles groundstate.
We exemplarily choose the $2_0\leftrightarrow 3_0$ resonance where, as seen in Fig.~\ref{fig2}(c), $F_{nv}=5/3$, but the results apply to other resonances as well. 
To this extent, let us observe that since the total Hamiltonian $H$ conserves particles, energy and spin, the stationary density matrix $\rho^\infty$ has a block diagonal structure, with blocks $\rho^{NSS_z}(E)$ of definite $N$, $S$, $S_z$, and $E$ \cite{Konig1996}. Due to the equivalence of the configurations with different $S_z$ for the dynamics, we introduce the matrices $\rho^{N}_{L_zL'_z}(E):=\sum_{S_z}(\rho^{N SS_z}(E))_{L_zL'_z}$. For example, since there exists only one configuration for the $N=2$ groundstate, $\rho^2$ is a number. On the other hand, $\rho^3(E_{3_0})$ is the $2\times 2$ matrix associated to the 3--particles groundstate quadruplet.
By using the Wigner-Eckart theorem \cite{Messiah1961} to calculate matrix elements of the operators, $d^\dagger_{\alpha\sigma}$ and $d_{\alpha\sigma}$, between states of different particle number and spin, and summing over $\sigma $, Eq.~(\ref{eq:liouvillian}) yields for the case of unidirectional transport near the $2_0 \leftrightarrow 3_0$ resonance
\begin{equation}
	0
	=
	-\frac{i}{\hbar} \left[H_{\mathrm{LS}},\rho^{3}\right]
	+ 2
	\Gamma\mathcal{ R}_L\rho^2
	-
	\frac{\Gamma}{2}\left\{ \mathcal{R}_R,\rho^{3}\right\}
	,
	\label{eq:56_resonance}
\end{equation}
which, together with $\mathrm{Tr}_\mathrm{TQD}\{\rho\}=1$, fully determines $\rho^{2}(E_{2_0})$ and $\rho^3(E_{3_0})$.
The Lamb shift Hamiltonian can be cast, following A.~Donarini et al. \cite{Donarini2009}, into the form $H_\mathrm{LS} = \hbar\sum_\alpha \omega_\alpha \mathcal{R}_\alpha$.
The precession frequencies $\omega_\alpha$ account for virtual transitions from the $3$--particles groundstates to the states with $2$ and $4$ particles
%the block $\rho^{N}(E^*)$ with spin $S$ are independent of $S_z$ ($\omega_{\alpha,S_z}=\omega_\alpha$).
and are independent of $S_z$. We find 
\begin{align}
\label{eq:precession_frequencies}
	&\omega_{\alpha}
	=
	\frac{\Gamma_{0}}{2\pi}
	\sum\limits_{\tau,E}
	p_\alpha\left(E-E_{3_0}\right)
	\times
	\\ \nonumber
	&\langle 3,E_{3_0};\frac{1}{2},S_z,L_z \vert
	d_{0\tau}
	\mathcal{P}_{4,E}
	d_{0\tau}^\dagger
	\vert 3,E_{3_0};\frac{1}{2},S_z,-L_z \rangle
	\\ \nonumber
	&+
	p_\alpha\left(E_{3_0}-E\right)
	\times
	\\ \nonumber
	&\langle 3,E_{3_0};\frac{1}{2},S_z,L_z \vert
	d_{0\tau}^\dagger
	\mathcal{P}_{2,E}
	d_{0\tau}
	\vert 3,E_{3_0};\frac{1}{2},S_z,-L_z \rangle
	,
\end{align}
where $\mathcal{P}_{NE}=\sum_{S_z,L_z}\vert N, E; S, S_z, L_z \rangle\langle N, E; S, S_z, L_z \vert$ is the projector on the $N$-particle level with energy $E$ and spin $S$. We defined the function $p_\alpha\left(\Delta E\right) = -\operatorname{Re} \psi\left[1/2+i(\Delta E - \mu_\alpha)/(2\pi k_\mathrm{B}T) \right]$ where $T$ is the temperature, $\psi$ the digamma function and $\mu_\alpha$ the chemical potential of lead $\alpha$.
The matrices $\mathcal{R}_\alpha$ have in the angular momentum basis the form $({\mathcal{R}}_\alpha)_{\ell\ell'}= e^{i\alpha(\ell-\ell')2\pi/3}$, $\ell,\ell'=\pm 1$.
We defined $\Gamma=a \Gamma_0$ where $a=\frac{1}{2}\sum_\sigma \vert \langle 3, E_{3_0}; \frac{1}{2} , -\sigma, 1 \vert d_{\alpha \sigma}^\dagger \vert 2_0 \rangle \vert^2$.
Notice that one cannot diagonalize  $\mathcal{R}_L$ and $\mathcal{R}_R$ simultaneuously. In the basis spanned by $\vert 3,DS \rangle$ and $\vert 3, CS  \rangle$ we get 
\begin{equation}
	\mathcal{R}_R
	=
	\begin{pmatrix}
		0 & 0 \\
		0 & 2
	\end{pmatrix},
	\qquad
	\mathcal{R}_L
	=
	\frac{1}{2}
	\begin{pmatrix}
		3 & -i\sqrt{3} \\
		i\sqrt{3} & 1
	\end{pmatrix}
	, 
	\label{eq:rate_matrices}
\end{equation}
and $\rho^3(E_{3_0}) = p(I+ \mathbf{n} \cdot \boldsymbol{\sigma})/2$, where $\boldsymbol{n}$ is the Bloch vector corresponding to the orbitally degenerate state, $\boldsymbol{\sigma}$ is the vector of Pauli matrices, $p=\rho_{dd}+\rho_{cc}$, and the decoupled state points along the $z$--axis.
Neglecting the Lamb shift term, the matrix $\rho^3(E_{3_0})$ is diagonal, with elements $\rho^{dd}=1$, $\rho^{cc}=0$ at deep interference blockade.
The diagonal elements $0$, $2$ and $3/2$, $1/2$ of $\mathcal R_\alpha$ correspond to the weights $R_R^s$, $R_R^f$, and $R_L^s$, $R_L^f$, respectively, entering Eq.~(\ref{eq:fanob}). 
Notice that this yields the counterintuitive result $R_L^f\neq R_L^s$.  
Using these values  we find $F_{nv}=5/3$.
\begin{figure}[t]
	\centering
	\includegraphics[width=\columnwidth]{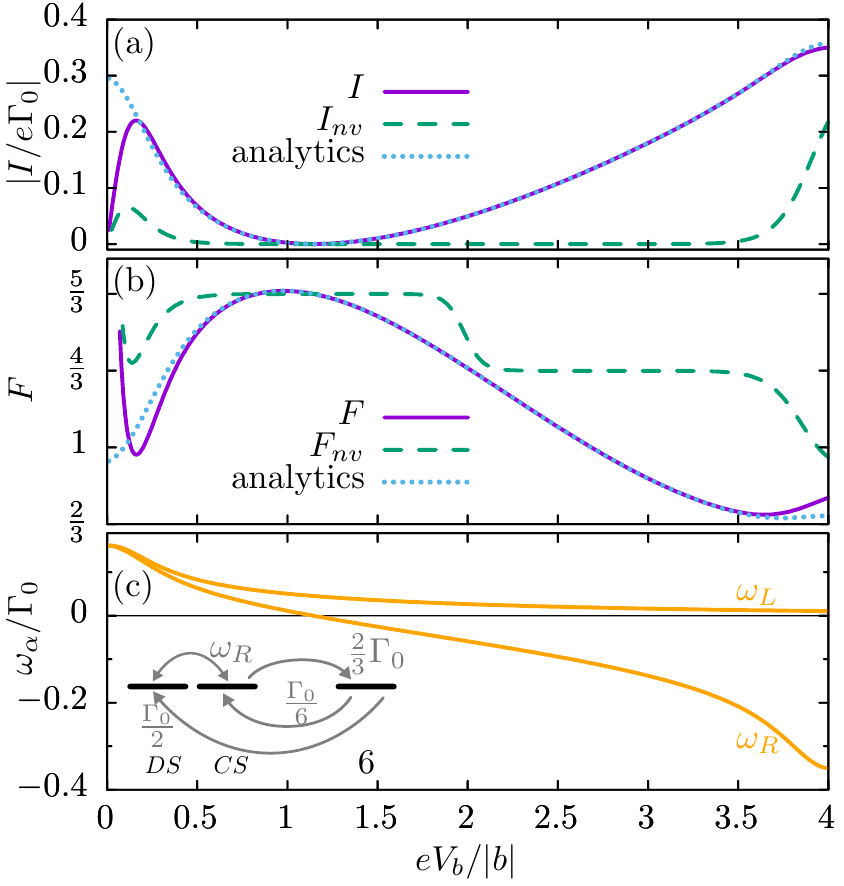}
	\caption{Bias trace of (a) current, (b) Fano factor and (c) precession frequencies $\omega_{L,R}$ at $\xi=-7.5\vert b \vert$, corresponding to the green dashed line in Fig.~\ref{fig2}. The numerical data well agree with analytical expressions from the text. Inset: The population of the coupled, $(CS)$, and dark, $(DS)$, states  is affected by effective filling rates and by a precessional dynamics with frequency $\omega_R$.
	}
	\label{fig3}
\end{figure}
So far the effect of the Lamb shift Hamiltonian $H_{LS}$ has been neglected. An analytical treatment of the precessional dynamics is possible in the parameter region involving the $N=5$ and $N=6$ groundstates as discussed below.

\section{Interference blockade at the $5_0\leftrightarrow 6_0$ resonance}
The Lamb shift term describes a precession of the Bloch vector $\boldsymbol{n}$ around an axis set by the matrices $\mathcal{R}_\alpha$. The populations of the coupled and of the dark state are thus affected by partially coherent gain and loss, and the blockade is perfect only when $\omega_L=0$.
We choose the $5_0\leftrightarrow 6_0$ resonance at $\xi=-7.5\vert b \vert$, indicated by a green dashed line in Fig.~\ref{fig2}, to study the effect of this precession.

\subsection{Hole transport}
The dynamics between the $N=5$ and $N=6$ groundstates is easily described in terms of hole transport. Then, Eqs.~(\ref{eq:fanob}), (\ref{eq:56_resonance}) and (\ref{eq:rate_matrices}) apply upon exchange of $L \leftrightarrow R$ together with $3\to 5$, $2\to 6$, which yields
\begin{equation}
\begin{split}
	\label{eq:master_equation}
	0
	&=
	-\frac{i}{\hbar} \left[H_{\mathrm{LS}},\rho^{5}\right]
	+ 2
	\Gamma\mathcal{R}_R\rho^6
	-
	\frac{\Gamma}{2}\left\{ \mathcal{R}_L,\rho^{5}\right\}
	, \\
	0
	&=
	\Gamma \mathrm{Tr_{TQD}}\left(\mathcal{R}_L\rho^{5}\right) - 4 \Gamma \rho^{6}
	,
\end{split}
\end{equation}
where we use $a=\frac{1}{2}\sum_\sigma \vert \langle 5, E_{5_0}; \frac{1}{2} , -\sigma, 1 \vert d_{\alpha \sigma} \vert 6 \rangle \vert^2=\frac{1}{3}$ and $\Gamma=a\Gamma_0$.

Bias traces of currents and Fano factors at the gate voltage corresponding to $\xi=-7.5\vert b\vert$ are shown in Fig.~\ref{fig3}(a) and \ref{fig3}(b).
The current $I_{nv}$, which does not account for the Lamb shifts, is exponentially suppressed in the voltage range $0.5<eV_\mathrm{b}/\vert b\vert < 3.5$. The associated Fano factor takes the values $F_{nv}=5/3$ at low bias, and $F_{nv}=4/3$ above $eV_\mathrm{b}/\vert b\vert= 2$, when transitions from $5_0$ to the $4$--particle groundstates $\{\vert 4, E_{4_0}; 1, S_z, 0 \rangle\}$ dominate the bottleneck process for transport, as shown in Appendix \ref{sec:A-4/3}.
Virtual transitions modify this picture: the current $I$ (Fano factor $F$) varies with bias voltage and has a minimum (maximum) at $\omega_{R} \approx 0$.
In the following we investigate how the Lamb shift Hamiltonian affects the dark state of the $5$--particles groundstates and the resulting Fano factor.

\subsection{Interference dynamics}
The precession frequencies $\omega_\alpha$ from Eq.~(\ref{eq:precession_frequencies}) account for virtual transitions from the $5$--particles groundstates $\{|5, E_{5_0}; 1/2, \pm 1/2, \pm 1 \rangle\}$ to the state $\vert 6_0 \rangle \equiv \vert 6, E_6; 0, 0, 0 \rangle$ and to levels with energies $E_{4_{0,1,2,3}}$. 
Their bias dependence is shown in Fig. \ref{fig3}(c). 
The stationary density matrix in the ordering $\rho^{dd},\rho^{cc},\rho^6,\rho^{dc},\rho^{cd}$ obtained as solution to Eq.~(\ref{eq:master_equation}) is
\begin{equation}
	\rho^\infty=
	\frac{1}{D}
	\begin{pmatrix}
		D-3\omega_R^2 \\
		2\omega_R^2 \\
		\omega_R^2 \\
		-\sqrt{3}\omega_R(\Gamma-i2(\omega_L-\omega_R)) \\
		-\sqrt{3}\omega_R(\Gamma+i2(\omega_L-\omega_R))
	\end{pmatrix}
	,
\end{equation}
with $D=2\Gamma^2+8\omega_L^2-12\omega_L\omega_R+9\omega_R^2$.
The corresponding current is $I=-e4\Gamma\omega_R^2/3D$.
For $\omega_R\rightarrow0$ the system gets quadratically stuck in the decoupled state and thus current is supressed.
The resulting Fano factor is
\begin{align}
	F
	=&
	\frac{
		16 \omega_R^2 \left(2 \Gamma^2+53 \omega_L^2\right)
		-176 \omega_L  \omega_R  \left(\Gamma^2+4 \omega_L^2\right)
	}{3 D^2}
	\nonumber \\ 
	&+
	\frac{
		20 \left(\Gamma^2+4 \omega_L^2\right)^2-576 \omega_L  \omega_R ^3+195 \omega_R ^4
	}{3 D^2}
	,
\end{align}
which to lowest order in $\omega_R$ is $F=5/3+16\omega_L\omega_R/(\Gamma^2+4\omega_L^2)$.
Therefore, the limit of $F=5/3$ is recovered at complete blockade. Since $\omega_L>0$, the Fano factor is not maximal at $\omega_R=0$ but instead at a little lower bias voltage.

\subsection{Robustness against perturbations}
\begin{figure*}[t]
	\includegraphics{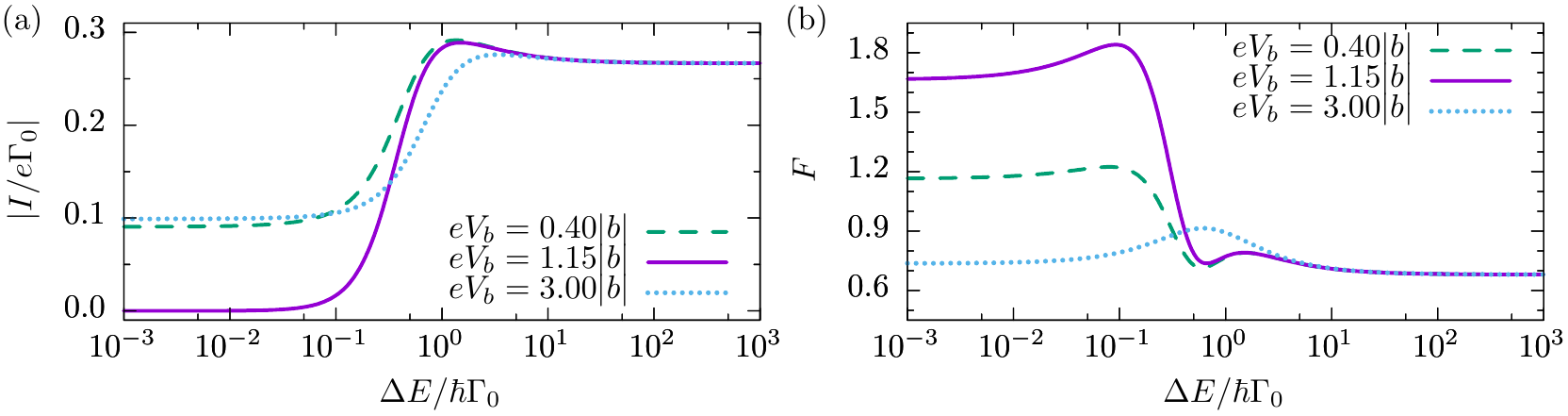}
	\caption{(a) Current and (b) Fano factor as a function of detuning at $V_\mathrm{g}=-7.5\vert b\vert$ for different bias voltages. In the limit of large detuning where $\Delta E \gg \hbar \Gamma_0$ the current is $I=-e4\Gamma_0/15$ and the Fano factor $F=17/25$. For vanishing detuning $F=5/3$ is recovered at full blockade ($eV_\mathrm{b}\approx 1.15\vert b \vert$).
	}
	\label{fig4}
\end{figure*}

To check the influence of a weak perturbation which lifts degeneracies, we restrict ourself again to the $5_0\leftrightarrow 6$ resonance. Then this perturbation changes the onsite energies of the orbitally degenerate $\vert 5_0\rangle$ states and is given by a Hamiltonian $H_\Delta=-\Delta E \sigma_z/2$ which, rotated to the coupled and decoupled basis, takes the form
\begin{equation}
	H_\Delta
	=
	\frac{1}{2}
	\begin{pmatrix}
		0 & \Delta E \\
		\Delta E & 0
	\end{pmatrix}.
\end{equation}
If one assumes weak coupling to the leads, $\Gamma\ll k_BT$ and in addition also that $\Delta E \ll k_BT$, the dissipative part of the unidirectional master equation is unaffected by the changes. Therefore, the master equation ~(\ref{eq:master_equation}) holds with the substitution $H_\mathrm{LS}\to H_\mathrm{LS}+H_\Delta$.
The resulting current and Fano factor at the $5_0\leftrightarrow 6$ resonance is shown as a function of the detuning in Fig.~\ref{fig4}.
The results of this paper are robust for a perturbation strength up to the order of $\Delta E \lesssim 0.01 \hbar \Gamma_0$. In the limit of large detuning where $k_BT \gg \Delta E \gg \hbar \Gamma_0$ the current is $I=-e4\Gamma_0/15$ and the Fano factor $F=17/25$. This Fano factor can be explained by the one for a single resonant level $F	= \left( R_L^2+R_R^2\right)/\left(R_L+R_R\right)^2$, where, due to the 4--fold degeneracy of $\vert 5_0 \rangle$, $R_L=4R_R$. Notice that $F=5/9$ at the right side of the stability diagram \cite{Thielmann2003} in Fig. 2(c) can be obtained with $R_L=2R_R$.

\section{Conclusions}
Using a full counting statistics approach in Liouville space we obtained the Fano stability diagram of a C$_{3v}$ symmetric TQD.
In the region of current suppression the Fano factor helps unraveling the underlying blocking mechanisms.
Poissonian statistics suggests ''classical'' Coulomb blockade, whereas super-Possonian noise points to the presence of fast and slow channels, the latter including dark states.
A population redistribution between dark and coupled states, induced by virtual excitations, results in a nontrivial bias dependence of the Fano factor.
The value attained by the Fano factor at specific gate and bias voltages further reveals the internal structure of the dark states.

This work is focused on the impact of the C$_{3v}$ symmetry on noise and thus restriction to sequential tunneling already yields interesting results. However, cotunneling contributions\cite{Weymann2011, Wrzesniewski2015} might additionally influence the noise features and should be further investigated for a C$_{3v}$ symmetric setup.

\begin{acknowledgments}
The authors acknowledge financial support by the Deutsche Forschungsgemeinschaft via GRK~1570 and SFB~689.
\end{acknowledgments}

%%%%%%%%%%%%%%%%%%%%%%%%%%%%%%%%%%%%%%%%%%%%%%%%%%%%%%%%%%%%%%%%%%%%%%%
\appendix
%%%%%%%%%%%%%%%%%%%%%%%%%%%%%%%%%%%%%%%%%%%%%%%%%%%%%%%%%%%%%%%%%%%%%%%

\section{Many--body spectrum and eigenfunctions of a symmetric TQD}
\label{sec:A-spectrum}
All eigenvectors and the associated eigenvalues in the occupation number basis of the TQD Hamiltonian Eq. (\ref{eq:hamiltonian}) are reported in Table~\ref{tab1} and \ref{tab2}. We also included all quantum numbers necessary to classify these states in terms of total particle number $N$, energy $E$, total spin $S^2$ and projections $S_z$ and $L_z$ of the total spin and angular momentum. The corresponding many-body state is denoted by $\vert N, E; S, S_z, L_z \rangle$.

We notice that a classification of many-body states using the angular momentum quantum number $L_z$ has been proposed by T. Kostyrko et al. \cite{Kostyrko2009} for the case of a symmetric triangular dot with intrasite repulsion only (i.e. $U\neq 0$,  $V=0$). Our analysis with $U\neq 0$, $V\neq 0$ thus generalizes that work and recovers the results reported by M. Korkusinski et al. \cite{Korkusinski2007}, where a localized representation is used to discuss topological Hund rules and derive effective low energy spin Hamiltonians.
For finite onsite and intersite interactions $U$ and $V$, the composition of the eigenstates is the result of a complex interplay between Pauli statistics and Coulomb repulsion, and we refer to the review by C.-Y. Hsieh et al. \cite{Hsieh2012} for useful insights. For example, for double occupancy of the TQD, the configurations with $S=1$ corresponds to excited states with singly occupied dots, due to Pauli principle. The configurations with $S=0$, however, contain both doubly occupied and singly occupied dots, with weight determined by the difference $U-V$.
For  $U=V$, the groundstate is in the occupation number representation the singlet $\vert 100, 100\rangle$, with equal weights on single and doubly occupied sites, as seen in Table~\ref{tab1}. The splitting between the sextuplet of excited states and the groundstate singlet is
dominated by the hopping energy with a correction given by super-exchange processes due to the doubly occupied singlet configurations \cite{Korkusinski2007}. For a TQD with occupancy $N=4$ (i.e. with two-holes), the groundstate is always a triplet if $b< 0$, as in our work. 
Finally, of relevance for the discussion in the main part of the manuscript, the 3--particles and 5--particles groundstates are a quadruplet due to orbital and spin degeneracy, while the associated first excited states are only spin degenerate.

%% Define lengthy expressions to make table compact!
%%%%%%%%%%%%%%%%%%%%%%%%%%%%%%%%%%%%%%%%%%%%%%%%%%%%%%%%%%%%%%%%%%%%%%%%%%%%%%%%%%%%%%%%%%%%%%%%%%%%%%%%%%%%%%%%%%%%%%%%%%%
% Eigenenergies
%%%%%%%%%%%%%%%%%%%%%%%%%%%%%%%%%%%%%%%%%%%%%%%%%%%%%%%%%%%%%%%%%%%%%%%%%%%%%%%%%%%%%%%%%%%%%%%%%%%%%%%%%%%%%%%%%%%%%%%%%%%
\def \EO {$E_{0}=0$}

\def \EIxO {$E_{1_0}=\xi-\frac{U}{2} - 2V + 2b$}
\def \EIxI {$E_{1_1}=\xi -\frac{U}{2} - 2V -b$}

\def \EIIxO {$E_{2_0}=2\xi - U - 3V +b +\frac{U-V}{2}-s_{-2}$}
\def \EIIxI {$E_{2_1}=2\xi- U - 3V +b$}
\def \EIIxII {$E_{2_2}=2\xi - U - 3V -\frac{b}{2}+\frac{U-V}{2}-s_1$}
\def \EIIxIII {$E_{2_3}=2\xi - U - 3V -2b$    }
\def \EIIxIV {$E_{2_4}=2\xi - U - 3V -\frac{b}{2} +\frac{U-V}{2}+s_1$}
\def \EIIxV {$E_{2_5}=2\xi+b - U - 3V +\frac{U-V}{2}+s_{-2}$}

\def \EIIIxO {$E_{3_0}=3\xi - \frac{3}{2}U -
3V+\frac{2}{3}\left(U-V\right)\left[1-\lambda_0/(2\vert a \vert)\right]$}
\def \EIIIxI {$E_{3_1}=3\xi- \frac{3}{2}U-3V$}
\def \EIIIxII {$E_{3_2}=3\xi-
\frac{3}{2}U-3V+\frac{2}{3}\left(U-V\right)\left[1-\lambda_1/(2\vert a
\vert)\right]$}
\def \EIIIxIII {$E_{3_3}=3\xi- \frac{3}{2}U-3V+\left(U-V\right)$}
\def \EIIIxIV {$E_{3_4}=3\xi- \frac{3}{2}U-3V
+\frac{2}{3}\left(U-V\right)\left[1-\lambda_{-1}/(2\vert a \vert)\right]$}

\def \EIVxO {$E_{4_0}=4\xi+2b - U - 3V$}
\def \EIVxI {$E_{4_1}=4\xi  - U - 3V +\frac{b}{2}+\frac{U-V}{2}-s_{-1}$}
\def \EIVxII {$E_{4_2}=4\xi - U - 3V-b+\frac{U-V}{2}-s_2$}
\def \EIVxIII {$E_{4_3}=4\xi  - U - 3V-b$}
\def \EIVxIV {$E_{4_4}=4\xi  - U - 3V +\frac{b}{2}+\frac{U-V}{2}+s_{-1}$}
\def \EIVxV {$E_{4_5}=4\xi  - U - 3V-b +\frac{U-V}{2}+s_2$}

\def \EVxO {$E_{5_0}=5\xi- \frac{U}{2} - 2V+b$}
\def \EVxI {$E_{5_1}=5\xi - \frac{U}{2}-2V -2b$}

\def \EVI {$E_{6}=6\xi$}
%%%%%%%%%%%%%%%%%%%%%%%%%%%%%%%%%%%%%%%%%%%%%%%%%%%%%%%%%%%%%%%%%%%%%%%%%%%%%%%%%%%%%%%%%%%%%%%%%%%%%%%%%%%%%%%%%%%%%%%%%%%
% Eigenstates
%%%%%%%%%%%%%%%%%%%%%%%%%%%%%%%%%%%%%%%%%%%%%%%%%%%%%%%%%%%%%%%%%%%%%%%%%%%%%%%%%%%%%%%%%%%%%%%%%%%%%%%%%%%%%%%%%%%%%%%%%%%
\def \SO {$\vert 000,000\rangle$}

\def \SIxOxII {$\vert 100,000\rangle$}
\def \SIxOxI {$\vert 000,100\rangle$}

\def \SIxIxIV {$\vert 010,000\rangle$}
\def \SIxIxII {$\vert 000,010\rangle$}
\def \SIxIxIII {$\vert 001,000\rangle$}
\def \SIxIxI {$\vert 000,001\rangle$}

\def \SIIxO {$\cos(\phi_{-2})\vert 100,100\rangle-\sin(\phi_{-2})\frac{1}{\sqrt{2}}\left(\vert 010,001 \rangle + \vert 001,010 \rangle \right)$}

\def \SIIxIxV {$\vert 101,000 \rangle$}
\def \SIIxIxIII {$\frac{1}{\sqrt{2}}\left(\vert 100,001 \rangle-\vert 001,100 \rangle \right)$}
\def \SIIxIxI {$\vert 000,101 \rangle$}
\def \SIIxIxVI {$\vert 110,000 \rangle$}
\def \SIIxIxIV {$\frac{1}{\sqrt{2}}\left(\vert 100,010 \rangle-\vert 010,100 \rangle \right)$}
\def \SIIxIxII {$\vert 000,110 \rangle$}

\def \SIIxIIxI {$\cos(\phi_1)\vert 010,010 \rangle-\sin(\phi_1)\frac{1}{\sqrt{2}}\left(\vert 100,001 \rangle + \vert 001,100 \rangle \right)$}
\def \SIIxIIxII {$\cos(\phi_1)\vert 001,001 \rangle-\sin(\phi_1)\frac{1}{\sqrt{2}}\left(\vert 100,010 \rangle + \vert 010,100 \rangle \right)$}

\def \SIIxIIIxIII {$\vert 011,000 \rangle$}
\def \SIIxIIIxII {$\frac{1}{\sqrt{2}}\left(\vert 010,001 \rangle-\vert 001,010 \rangle \right)$}
\def \SIIxIIIxI {$\vert 000,011 \rangle$}

\def \SIIxIVxI {$\sin(\phi_1)\vert 010,010 \rangle+\cos(\phi_1)\frac{1}{\sqrt{2}}\left(\vert 100,001 \rangle + \vert 001,100 \rangle \right)$}
\def \SIIxIVxII {$\sin(\phi_1)\vert 001,001 \rangle+\cos(\phi_1)\frac{1}{\sqrt{2}}\left(\vert 100,010 \rangle + \vert 010,100 \rangle \right)$}

\def \SIIxV {$\sin(\phi_{-2})\vert 100,100 \rangle+\cos(\phi_{-2})\frac{1}{\sqrt{2}}\left(\vert 010,001 \rangle + \vert 001,010 \rangle \right)$}

\def \SIIIxOxIV {$ v_{0,1}\vert 110,100 \rangle - v_{0,0}\vert 101,001 \rangle + v_{0,-1}\vert 011,010 \rangle$}
\def \SIIIxOxII {$ v_{0,1}\vert 100,110\rangle + v_{0,0}\vert 001,101 \rangle - v_{0,-1}\vert 010,011 \rangle$}
\def \SIIIxOxIII {$ v_{0,1}\vert 101,100\rangle - v_{0,0}\vert 110,010 \rangle - v_{0,-1}\vert 011,001 \rangle$}
\def \SIIIxOxI {$ v_{0,1}\vert 100,101\rangle - v_{0,0}\vert 010,110 \rangle - v_{0,-1}\vert 001,011 \rangle$}

\def \SIIIxIxIV {$\vert 111,000 \rangle$}
\def \SIIIxIxIII {$\frac{1}{\sqrt{3}}\left(\vert 011,100 \rangle-\vert 101,010 \rangle+\vert 110,001 \rangle \right)$}
\def \SIIIxIxII {$\frac{1}{\sqrt{3}}\left(\vert 001,110\rangle-\vert 010,101 \rangle+\vert 100,011 \rangle \right)$}
\def \SIIIxIxI {$\vert 000,111 \rangle$}

\def \SIIIxIIxI {$ v_{1,1}\vert 110,100 \rangle - v_{1,0}\vert 101,001 \rangle + v_{1,-1}\vert 011,010 \rangle$}
\def \SIIIxIIxII {$ v_{1,1}\vert 100,110 \rangle + v_{1,0}\vert 001,101 \rangle - v_{1,-1}\vert 010,011 \rangle$}
\def \SIIIxIIxIII {$ v_{1,1}\vert 101,100 \rangle - v_{1,0}\vert 110,010 \rangle - v_{1,-1}\vert 011,001 \rangle$}
\def \SIIIxIIxIV {$ v_{1,1}\vert 100,101 \rangle - v_{1,0}\vert 010,110 \rangle - v_{1,-1}\vert 001,011 \rangle$}

\def \SIIIxIIIxIII {$\frac{1}{\sqrt{6}}\left(\vert 110,001 \rangle+2\vert 101,010 \rangle+\vert 011,100 \rangle \right)$}
\def \SIIIxIIIxII {$\frac{1}{\sqrt{6}}\left(\vert 001,110 \rangle+2\vert 010,101 \rangle+\vert 100,011 \rangle \right)$}
\def \SIIIxIIIxIV {$\frac{1}{\sqrt{2}}\left(\vert 110,001 \rangle-\vert 011,100 \rangle \right)$}
\def \SIIIxIIIxI {$\frac{1}{\sqrt{2}}\left(\vert 001,110 \rangle-\vert 100,011 \rangle \right)$}

\def \SIIIxIVxIV {$ v_{-1,1}\vert 110,100 \rangle - v_{-1,0}\vert 101,001 \rangle + v_{-1,-1}\vert 011,010 \rangle$}
\def \SIIIxIVxII {$ v_{-1,1}\vert 100,110 \rangle + v_{-1,0}\vert 001,101 \rangle - v_{-1,-1}\vert 010,011 \rangle$}
\def \SIIIxIVxIII {$ v_{-1,1}\vert 101,100 \rangle - v_{-1,0}\vert 110,010 \rangle - v_{-1,-1}\vert 011,001 \rangle$}
\def \SIIIxIVxI {$ v_{-1,1}\vert 100,101 \rangle - v_{-1,0}\vert 010,110 \rangle - v_{-1,-1}\vert 001,011 \rangle$}

\def \SIVxOxIII {$\vert 111,100 \rangle$}
\def \SIVxOxII {$\frac{1}{\sqrt{2}}\left(\vert 101,110 \rangle-\vert 110,101 \rangle \right)$}
\def \SIVxOxI {$\vert 100,111 \rangle$}

\def \SIVxIxII {$\cos(\phi_{-1})\vert 101,101 \rangle-\sin(\phi_{-1})\frac{1}{\sqrt{2}}\left(\vert 011,110 \rangle + \vert 110,011 \rangle \right)$}
\def \SIVxIxI {$\cos(\phi_{-1})\vert 110,110 \rangle+\sin(\phi_{-1})\frac{1}{\sqrt{2}}\left(\vert 011,101 \rangle + \vert 101,011 \rangle \right)$}

\def \SIVxII {$\cos(\phi_2)\vert 011,011\rangle+\sin(\phi_2)\frac{1}{\sqrt{2}}\left(\vert 101,110 \rangle + \vert 110,101 \rangle \right)$}

\def \SIVxIIIxVI {$\vert 111,010 \rangle$}
\def \SIVxIIIxIV {$\frac{1}{\sqrt{2}}\left(\vert 011,110 \rangle-\vert 110,011 \rangle \right)$}
\def \SIVxIIIxII {$\vert 010,111 \rangle$}
\def \SIVxIIIxV {$\vert 111,001 \rangle$}
\def \SIVxIIIxIII {$\frac{1}{\sqrt{2}}\left(\vert 011,101 \rangle-\vert 101,011 \rangle \right)$}
\def \SIVxIIIxI {$\vert 001,111 \rangle$}

\def \SIVxIVxII {$\sin(\phi_{-1})\vert 101,101 \rangle+\cos(\phi_{-1})\frac{1}{\sqrt{2}}\left(\vert 011,110 \rangle + \vert 110,011 \rangle \right)$}
\def \SIVxIVxI {$\sin(\phi_{-1})\vert 110,110 \rangle-\cos(\phi_{-1})\frac{1}{\sqrt{2}}\left(\vert 011,101 \rangle + \vert 101,011 \rangle \right)$}

\def \SIVxV {$-\sin(\phi_2)\vert 011,011 \rangle+\cos(\phi_2)\frac{1}{\sqrt{2}}\left(\vert 101,110 \rangle + \vert 110,101 \rangle \right)$}

\def \SVxOxIV {$\vert 111,110 \rangle$}
\def \SVxOxII {$\vert 110,111 \rangle$}
\def \SVxOxIII {$\vert 111,101 \rangle$}
\def \SVxOxI {$\vert 101,111 \rangle$}

\def \SVxIxII {$\vert 111,011 \rangle$}
\def \SVxIxI {$\vert 011,111 \rangle$}

\def \SVI {$\vert 111,111 \rangle$}

%%%%%%%%%%%%%%%%%%%%%%%%%%%%%%%%%%%%%%%%%%%%%%%%%%%%%%%%%%%%%%%%%%%%%%%%%%%%%%%%%%%%%%%%%%%%%%%%%%%%%%%%%%%%%%%%%%%%%%%%%%%
% Tables
%%%%%%%%%%%%%%%%%%%%%%%%%%%%%%%%%%%%%%%%%%%%%%%%%%%%%%%%%%%%%%%%%%%%%%%%%%%%%%%%%%%%%%%%%%%%%%%%%%%%%%%%%%%%%%%%%%%%%%%%%%%

\begin{table*}[ht]
\centering
\begin{tabular}{|c|l|c|c|c|l|}
\hline
$N$						& Eigenenergy				& $S$								& $S_z$								& $L_z$					& Eigenstate in the basis $\{\vert n_{0\uparrow},n_{1\uparrow},n_{-1\uparrow};n_{0\downarrow},n_{1\downarrow},n_{-1\downarrow} \rangle\}$ 	\\
\hline
\hline

$0$						& \EO						& $0$								& $0$								& $0$					& \SO			\\
\hline

\multirow{6}{*}{$1$}	& \multirow{2}{*}{\EIxO}	& \multirow{2}{*}{$\frac{1}{2}$}	& $-\frac{1}{2}$					& \multirow{2}{*}{$0$}	& \SIxOxI		\\
\cline{4-4}\cline{6-6}
						&							& 									& $\frac{1}{2}$						& 						& \SIxOxII		\\
\cline{2-6}
						& \multirow{4}{*}{\EIxI}	& \multirow{4}{*}{$\frac{1}{2}$}	& \multirow{2}{*}{$-\frac{1}{2}$}	& $-1$					& \SIxIxI		\\
\cline{5-6}
						&							& 									& 									& $1$					& \SIxIxII		\\
\cline{4-6}
						&							& 									& \multirow{2}{*}{$\frac{1}{2}$}	& $-1$					& \SIxIxIII		\\
\cline{5-6}
						&							& 									&									& $1$					& \SIxIxIV		\\
\hline

\multirow{15}{*}{$2$}	& \EIIxO					& $0$								& $0$								& $0$					& \SIIxO		\\
\cline{2-6}
						& \multirow{6}{*}{\EIIxI}	& \multirow{6}{*}{$1$}				& \multirow{2}{*}{$-1$}				& $-1$					& \SIIxIxI	\\
\cline{5-6}
						& 							&									& 									& $1$					& \SIIxIxII		\\
\cline{4-6} 	
						& 							& 									& \multirow{2}{*}{$0$}				& $-1$					& \SIIxIxIII	\\
\cline{5-6} 	
						& 							& 									&									& $1$					& \SIIxIxIV		\\
\cline{4-6}
						& 							& 									& \multirow{2}{*}{$1$}				& $-1$					& \SIIxIxV		\\
\cline{5-6}
						& 							& 									&									& $1$					& \SIIxIxVI		\\
\cline{2-6}
						& \multirow{2}{*}{\EIIxII}	& \multirow{2}{*}{$0$}				& \multirow{2}{*}{$0$}				& $-1$					& \SIIxIIxI		\\
\cline{5-6}
						& 							& 									& 									& $1$					& \SIIxIIxII	\\
\cline{2-6}
						& \multirow{3}{*}{\EIIxIII}	& \multirow{3}{*}{$1$}				& $-1$								& \multirow{3}{*}{$0$}	& \SIIxIIIxI	\\
\cline{4-4}\cline{6-6}
						&							& 									& $0$								& 						& \SIIxIIIxII	\\
\cline{4-4}\cline{6-6}
						&							&									& $1$								& 						& \SIIxIIIxIII	\\
\cline{2-6} 	
						& \multirow{2}{*}{\EIIxIV}	& \multirow{2}{*}{$0$}				& \multirow{2}{*}{$0$}				& $-1$					& \SIIxIVxI		\\
\cline{5-6}
						& 							& 									& 									& $1$ 					& \SIIxIVxII	\\
\cline{2-6}
						& \EIIxV					& $0$								& $0$								& $0$ 					& \SIIxV		\\
\hline

\multirow{20}{*}{$3$}	& \multirow{4}{*}{\EIIIxO}	& \multirow{4}{*}{$\frac{1}{2}$}	& \multirow{2}{*}{$-\frac{1}{2}$}	& $-1$ 					& \SIIIxOxI		\\
\cline{5-6}
						& 							& 									& 									& $1$					& \SIIIxOxII	\\
\cline{4-6}
						& 							& 									& \multirow{2}{*}{$\frac{1}{2}$}	& $-1$ 					& \SIIIxOxIII	\\
\cline{5-6}
						& 							& 									& 									& $1$					& \SIIIxOxIV	\\
\cline{2-6}
						& \multirow{4}{*}{\EIIIxI}	& \multirow{4}{*}{$\frac{3}{2}$}	& $-\frac{3}{2}$					& \multirow{4}{*}{$0$}	& \SIIIxIxI		\\
\cline{4-4}\cline{6-6}
						& 							& 									& $-\frac{1}{2}$					& 						& \SIIIxIxII	\\
\cline{4-4}\cline{6-6}
						& 							& 									& $\frac{1}{2}$						& 						& \SIIIxIxIII	\\
\cline{4-4}\cline{6-6}
						& 							& 									& $\frac{3}{2}$						& 						& \SIIIxIxIV	\\
\cline{2-6}
						& \multirow{4}{*}{\EIIIxII}	& \multirow{4}{*}{$\frac{1}{2}$}	& \multirow{2}{*}{$-\frac{1}{2}$}	& $-1$ 					& \SIIIxIIxI	\\
\cline{5-6}
						& 							& 									& 									& $1$					& \SIIIxIIxII	\\
\cline{4-6}
						& 							& 									&\multirow{2}{*}{$\frac{1}{2}$}	& $-1$				& \SIIIxIIxIII	\\
\cline{5-6}
						& 							& 									& 									& $1$ 					& \SIIIxIIxIV	\\
\cline{2-6}
						& \multirow{4}{*}{\EIIIxIII}& \multirow{4}{*}{$\frac{1}{2}$}	& \multirow{2}{*}{$-\frac{1}{2}$}	& \multirow{4}{*}{$0$}	& \SIIIxIIIxI	\\
\cline{6-6}
						& 							& 									& 									& 						& \SIIIxIIIxII	\\
\cline{4-4}\cline{6-6}
						& 							& 									& \multirow{2}{*}{$\frac{1}{2}$}	& 						& \SIIIxIIIxIII	\\
\cline{6-6}
						& 							& 									& 									& 						& \SIIIxIIIxIV	\\
\cline{2-6}
						& \multirow{4}{*}{\EIIIxIV}	& \multirow{4}{*}{$\frac{1}{2}$}	& \multirow{2}{*}{$-\frac{1}{2}$}	& $-1$					& \SIIIxIVxI	\\
\cline{5-6}
						& 							& 									& 									& $1$					& \SIIIxIVxII	\\
\cline{4-6}
						& 							& 									& \multirow{2}{*}{$\frac{1}{2}$}	& $-1$					& \SIIIxIVxIII	\\
\cline{5-6}
						& 							& 									& 									& $1$					& \SIIIxIVxIV	\\
\hline
\end{tabular}
\caption{Eigenvalues and eigenstates of a C$_{3v}$ symmetric TQD Hamiltonian for occupation numbers $N=0-3$. Such eigenvectors are furthermore characterized by the spin quantum numbers $S$ and $S_z$, and by the orbital quantum number $L_z$. Their composition in the basis of the occupation number vectors
is provided in the rightmost column.  The ordering of the eigenergies depends on the TQD parameters $b, U$ and $V$. We chose $U=5|b|$, $V=2|b|$ and $b<0$. We defined $a=(U-V)/(9b)$, $\theta=\arccos\left(\left((3a^2)/(1+3a^2)\right)^{\frac{3}{2}}\right)/3$, $\lambda_\alpha=2\sqrt{(1+a^2)/3}\cos\left(\theta+\alpha\frac{2\pi}{3}\right)$, $v_{x,y}=(a-\lambda_x)\vert(a-\lambda_x)^2-1\vert/(a-\lambda_x-y)\sqrt{3(a-\lambda_x)^4+1}$, $s_x=\sqrt{9x^2b^2 + xb/2\left(U-V\right)+\left(U-V\right)^2}$ and $\phi_x=\frac{1}{2}\arctan\left(\frac{2\sqrt{2}(U-V)}{U-V+9xb}\right)$.}
\label{tab1}
\end{table*}

\begin{table*}[ht]
\centering
\begin{tabular}{|c|l|c|c|c|l|}
\hline
$N$						& Eigenenergy				& $S$								& $S_z$								& $L_z$					& Eigenstate in the basis $\{\vert n_{0\uparrow},n_{1\uparrow},n_{-1\uparrow};n_{0\downarrow},n_{1\downarrow},n_{-1\downarrow} \rangle\}$ 	\\
\hline
\hline

\multirow{15}{*}{$4$}	& \multirow{3}{*}{\EIVxO}	& \multirow{3}{*}{$1$}				& $-1$								& \multirow{3}{*}{$0$}	& \SIVxOxI		\\
\cline{4-4}\cline{6-6}
						&							& 									& $0$								& 						& \SIVxOxII		\\
\cline{4-4}\cline{6-6}
						&							& 									& $1$								& 						& \SIVxOxIII	\\
\cline{2-6}
						& \multirow{2}{*}{\EIVxI}	& \multirow{2}{*}{$0$}				& \multirow{2}{*}{$0$}				& $-1$					& \SIVxIxI		\\
\cline{5-6}
						& 							&									& 									& $1$					& \SIVxIxII		\\
\cline{2-6}
						& \EIVxII					& $0$								& $0$								& $0$					& \SIVxII		\\
\cline{2-6}
						& \multirow{6}{*}{\EIVxIII}	& \multirow{6}{*}{$1$}				& \multirow{2}{*}{$-1$}				& $-1$					& \SIVxIIIxI	\\
\cline{5-6}
						& 							& 									&									& $1$					& \SIVxIIIxII	\\
\cline{4-6}
						& 							& 									& \multirow{2}{*}{$0$}				& $-1$					& \SIVxIIIxIII	\\
\cline{5-6}
						& 							& 									& 									& $1$					& \SIVxIIIxIV	\\
\cline{4-6}
						& 							& 									& \multirow{2}{*}{$1$}				& $-1$					& \SIVxIIIxV	\\
\cline{5-6}
						& 							& 									& 									& $1$					& \SIVxIIIxVI	\\
\cline{2-6}
						& \multirow{2}{*}{\EIVxIV}	& \multirow{2}{*}{$0$}				& \multirow{2}{*}{$0$}				& $-1$					& \SIVxIVxI		\\
\cline{5-6}
						& 							& 									& 									& $1$					& \SIVxIVxII	\\
\cline{2-6}
						& \EIVxV					& $0$								& $0$								& $0$					& \SIVxV		\\
\hline

\multirow{6}{*}{$5$}	& \multirow{4}{*}{\EVxO}	& \multirow{4}{*}{$\frac{1}{2}$}	& \multirow{2}{*}{$-\frac{1}{2}$}	& $-1$					& \SVxOxI		\\
\cline{5-6}
						& 							&									& 									& $1$					& \SVxOxII		\\
\cline{4-6}
						& 							& 									& \multirow{2}{*}{$\frac{1}{2}$}	& $-1$					& \SVxOxIII		\\
\cline{5-6}
						& 							& 									& 									& $1$					& \SVxOxIV		\\
\cline{2-6}
						& \multirow{2}{*}{\EVxI}	& \multirow{2}{*}{$\frac{1}{2}$}	& $-\frac{1}{2}$					& \multirow{2}{*}{$0$}	& \SVxIxI		\\
\cline{4-4}\cline{6-6}
						& 							& 									& $\frac{1}{2}$						& 						& \SVxIxII		\\
\hline	
					
$6$						& \EVI						& $0$								& $0$								& $0$					& \SVI			\\
\hline
\end{tabular}
\caption{Eigenvalues and eigenstates of a C$_{3v}$ symmetric TQD for electron numbers $N=4$--$6$. The parameters and notations are chosen as in Table \ref{tab1}. The ordering is for $U=5|b|$, $V=2|b|$ and $b<0$. We defined $s_x=\sqrt{9x^2b^2 + xb/2\left(U-V\right)+\left(U-V\right)^2}$ and $\phi_x=\frac{1}{2}\arctan\left(\frac{2\sqrt{2}(U-V)}{U-V+9xb}\right)$.}
\label{tab2}
\end{table*}

\section{Current and Fano factor for a minimal model with slow and fast channels}
\label{sec:A-Fnv}
Let us consider a minimal system consisting of a slow and fast channel which for example can be a groundstate in Coulomb blockade (CB) plus an excited state in the bias window, or the coupled and decoupled states in the case of interference blockade (IB), as depicted in Figs. \ref{fig1}(b),(c) respectively. This system spends most of the time in the state corresponding to the slow channel and therefore the exponentially suppressed current is dominated by the bottleneck process of escaping this state.
The Liouvillian $\mathcal{L}$ or the tunneling Liouvillian $\mathcal{L}_t$ are super-operators whose matrix elements are obtained from their action on the reduced density matrix. In order to obtain a suitable representation of such Liouvillians, it is convenient to work in the Liouville space, where the density matrix elements are ordered in a vector.
Let us consider a situation of positive electrochemical potential, such that particle transport occurs from the left, ${L}$, to the right, ${R}$, lead. Then, far from resonance (e.g. inside the Coulomb blockade or interference blockade regions), we can approximate the Fermi functions of the fast channel to $1$ or $0$. In a generic situation where a state $\vert p \rangle$ has one less electrons as the fast $\vert f \rangle$ and slow $\vert s \rangle$ states, the Liouvillian in the basis $\{\vert p \rangle\langle p \vert, \vert f \rangle \langle f \vert, \vert s \rangle \langle s \vert \}$ takes the form
\begin{equation}
	\label{eq:liouvillian_at_blockade}
	\mathcal{L}_t
	%=
	%\mathcal{L}_t^f + \mathcal{L}_t^s
	=
	\begin{pmatrix}
		-\Gamma_L^f & \Gamma_R^f  & 0 \\
		\Gamma_L^f  & -\Gamma_R^f & 0 \\
		0           & 0           & 0
	\end{pmatrix} + \sum\limits_\alpha
	\begin{pmatrix}
		-\Gamma^s_\alpha f^+_\alpha & 0 & \Gamma_\alpha^sf^-_\alpha \\
		0                           & 0 & 0                         \\
		\Gamma^s_\alpha f^+_\alpha  & 0 & -\Gamma_\alpha^sf^-_\alpha
	\end{pmatrix}.
\end{equation}
The corresponding current superoperators for the right lead are
\begin{equation}
	\label{eq:current_operators_at_blockade}
	\mathcal{J}^+_R = \begin{pmatrix}
		0 & \Gamma_R^f & \Gamma_R^sf^-_R \\
		0 & 0 & 0 \\
		0 & 0 & 0
	\end{pmatrix}
	,\quad
	\mathcal{J}^-_R = \begin{pmatrix}
		0 & 0 & 0 \\
		0 & 0 & 0 \\
		\Gamma^s_R f^+_R & 0 & 0
	\end{pmatrix}.
\end{equation}
In our considerations, $\Gamma_\alpha^s f^-_\alpha$ gives the bottleneck for transport, therefore we use $f^+_\alpha=1-f^-_\alpha$ and expand to lowest order in $\Gamma_\alpha^s f^-_\alpha$ to obtain the current in the right lead
\begin{equation}
	\label{eq:current_at_blockade}
	I_R
	=
	-e \frac{\left( \Gamma^f_L+\Gamma^s_L \right) \Gamma^s_R f^-_R + \left( \Gamma^f_L-\Gamma^s_R \right) \Gamma^s_L f^-_L}{\Gamma^s_L + \Gamma^s_R}
	.
\end{equation}
In a CB situation with $f^-_R \gg f^-_L$, and for large asymmetry between the couplings to the leads, $\Gamma_L^s\gg\Gamma_R^s$, Eq.~(\ref{eq:current_at_blockade}) simplifies to $I_R=-e\Gamma^s_R f^-_R (1+\Gamma^f_L/\Gamma^s_L)$, in agreement with the findings by W.~Belzig et al. \cite{Belzig2005}.
In the case of IB where $\Gamma^s_R=0$, and the current becomes $I_R=-e\Gamma^f_Lf^-_L$.
Similar calculations as for the current yield the Fano factor
\begin{equation}
	\label{eq:fano_at_blockade}
	F_{nv}
	=
	1+\frac{2\Gamma^f_L}{\Gamma^s_L + \Gamma^s_R}
	.
\end{equation}
Notice that it always holds $F_{nv} \geq 1$.
The expression simplifies to $F_{nv}=1+2\Gamma_L^f/\Gamma_L^s$ for CB with large asymmetry \citep{Belzig2005} and IB. 
One can show that in a situation where the state $\vert p \rangle$ has one electrons less than $\vert f \rangle$ and $\vert s \rangle$ the Liouvillian and the current operators for the left lead are given again by Eqs. (\ref{eq:liouvillian_at_blockade}) and (\ref{eq:current_operators_at_blockade}), respectively, upon exchanging $L \leftrightarrow R$ and $f^+_\alpha \leftrightarrow f^-_\alpha$. This corresponds to hole transport. With the same exchanges the current and Fano factor are obtained from Eqs.~(\ref{eq:current_at_blockade}) and (\ref{eq:fano_at_blockade}), respectively.
Notice that this result includes also external asymmeties $\Gamma_L \neq \Gamma_R$.

\section{Liouvillian, current and Fano factor at the $5_0\leftrightarrow6$ resonance}
The far left part of the stability diagram shown in Fig. \ref{fig2} is dominated by $5\leftrightarrow 6$ particle transitions. As seen in Table \ref{tab2}, there exists only one configuration with $6$ electrons given by the state $\vert 6\rangle:=\vert 6,E_6;0,0,0\rangle$. On the other hand, when $5$ electrons populate the TQD a total of $6$ configurations are possible. In particular, for $b<0$ the groundstate is the quadruplet $\{\vert 5,E_{5_0};1/2,\pm 1/2,\pm 1\rangle\}$, due to both orbital and spin degeneracy. The first excited state is the doublet $\{\vert 5,E_{5_1};1/2,\pm 1/2,0\rangle\}$ and is only spin degenerate.
In this section we provide the explicit form of the tunneling Liouvillian $\mathcal{L}_t$ when the Fock space is restricted to the subspaces associated to the $5$-- and $6$--particles groundstates.
Such a Liouvillian determines the stationary reduced density matrices $\rho^{5}(E_{5_0})$ and $\rho^{6}$ when Lamb shifts are neglected. Since the $6$ particle groundstate is a singlet, $\rho^6$ is just a $1\times 1$ matrix. On the other hand, the calculation of $\rho^{5}(E_{5_0})$ involves first the evaluation of the matrix elements $\rho^{5\frac{1}{2}S_z}_{L_zL'_z}=\langle 5,E_{5_0};1/2,S_z,L_z\vert \rho^\infty\vert 5,E_{5_0};1/2,S_z,L'_z\rangle$, and then a summation over $S_z$: $\rho^5_{L_zL'_z}:=\sum_{S_z}\rho^{5 \frac{1}{2}S_z}_{L_zL_z}$. Hence, $\rho^5(E_{5_0})$ is a $2\times 2$ matrix in a basis spanned by the vectors $\vert 5_0^+\rangle$ and $\vert 5_0^-\rangle$, where $\pm$ refers to the associated values of the angular momentum. Notice that coherences between states with the same particle number but different angular momentum have to be considered.
We choose the basis
\{$\vert 5_0^+ \rangle \langle 5_0^+ \vert$,
$\vert 5_0^- \rangle \langle 5_0^- \vert$,
$\vert 6 \rangle \langle 6 \vert$,
$\vert 5_0^+ \rangle \langle 5_0^-\vert $,
$\vert 5_0^- \rangle \langle 5_0^+\vert $\}. Then the tunneling Liouvillian can be written as
\begin{widetext}
\begin{equation}
	\mathcal{L}_t
	=
	\frac{\Gamma_0}{3}
	\begin{pmatrix}
		 -f^+_R-f^+_L & 0 & 2(f^-_R+f^-_L) & \frac{1}{2}X & \frac{1}{2}X^*		\\
		 0 & -f^+_R-f^+_L & 2(f^-_R+f^-_L) & \frac{1}{2}X & \frac{1}{2}X^*		\\
		 f^+_R+f^+_L & f^+_R+f^+_L & -4(f^-_R+f^-_L) & -X & -X^*		 		\\
		 \frac{1}{2}X^* & \frac{1}{2}X^* & 2(1+X^*) & -f^+_R-f^+_L & 0		 	\\
		 \frac{1}{2}X & \frac{1}{2}X & 2(1+X) & 0 & -f^+_R-f^+_L
	\end{pmatrix},
\end{equation}
\end{widetext}
with the short notation $X=e^{-i\frac{2\pi}{3}}f^+_R+e^{i\frac{2\pi}{3}}f^+_L$.
Note that the phases $e^{\pm i\frac{2\pi }{3}}$ arise from changing from position into angular momentum basis.
The solution of the equation $\mathcal{L}_t\rho^\infty=0$, together with the constraint $\mathrm{Tr}_\mathrm{TQD}\{ \rho^\infty \}=1$ (with $\mathrm{Tr}_\mathrm{TQD}=(1,1,1,0,0)$), yields the stationary density matrix vector
\begin{equation}
	\rho^\infty
	=
	\frac{1}{D}
	\begin{pmatrix}
		f^+_Rf^-_L + f^+_Lf^-_R \\
		f^+_Rf^-_L + f^+_Lf^-_R \\
		f^+_Rf^+_L \\
		\frac{1}{2}(f^+_R-f^+_L)\left[\frac{f^+_R-f^+_L}{f^+_R+f^+_L}-i\sqrt{3}\right] \\
		-\frac{1}{2}(f^+_R-f^+_L)\left[\frac{f^+_R-f^+_L}{f^+_R+f^+_L}+i\sqrt{3}\right]
	\end{pmatrix},
\end{equation}
with $D=2(f^+_Rf^-_L + f^+_Lf^-_R)+f^+_Rf^+_L$.

The current operators for the right lead can be calculated along the same lines and read in the same basis
\begin{equation}
\begin{split}
	\mathcal{J}^+
	&=
	\frac{\Gamma_0}{3} f^-_R
	\begin{pmatrix}
		0 && 0 && 2 && 0 && 0 \\
		0 && 0 && 2 && 0 && 0 \\
		0 && 0 && 0 && 0 && 0 \\
		0 && 0 && -2e^{i\frac{2\pi}{3}} && 0 && 0 \\
		0 && 0 && -2e^{-i\frac{2\pi}{3}} && 0 && 0
	\end{pmatrix}
	,\\
	\mathcal{J}^-
	&=
	\frac{\Gamma_0}{3} f^+_R
	\begin{pmatrix}
		0 && 0 && 0 && 0 && 0 \\
		0 && 0 && 0 && 0 && 0 \\
		1 && 1 && 0 && -e^{-i\frac{2\pi}{3}} && -e^{i\frac{2\pi}{3}} \\
		0 && 0 && 0 && 0 && 0 \\
		0 && 0 && 0 && 0 && 0
	\end{pmatrix},
\end{split}
\end{equation}
which results in the current through the right lead
\begin{equation}
	I
	=
	-e
	\frac{\Gamma_0}{3}\frac{(f^+_L-f^+_R)f^+_Rf^+_L}{(f^+_R+f^+_L)\left[2(f^+_Rf^-_L+f^+_Lf^-_R)+f^+_Rf^+_L\right]},
\end{equation}
and the Fano factor
\begin{equation}
\begin{split}
	F_{nv}
	=&
	\frac{
		2 (3 f_L^-+1) (f_R^+)^2 \left(
			(f_L^+)^2+2 f_L^+ f_L^--2 f_L^- f_R^+
		\right)
	}{
		(f_L^++f_R^+) (f_L^+ f_R^-+f_L^++2 f_L^- f_R^+)^2
	}
	\\
	&+\frac{
		5 (f_L^+)^2+5 f_L^+ f_R^++3 (f_R^+)^2
	}{
		(f_L^++f_R^+)^2
	}
	+\frac{
		6 f_L^- f_R^+
	}{
		f_R^+-f_L^+
	}
	\\
	&-\frac{
		f_R^+ \left[
			(f_L^+)^2 (8 f_L^-+f_R^++4)
		\right]
	}{
		(f_L^++f_R^+)^2 (f_L^+ f_R^-+f_L^++2 f_L^- f_R^+)
	}
	\\
	&+\frac{
		f_R^+ \left[
			4 f_L^+ f_L^- f_R^++4 f_L^- (f_R^+)^2
		\right]
	}{
		(f_L^++f_R^+)^2 (f_L^+ f_R^-+f_L^++2 f_L^- f_R^+)
	}
\end{split}
\end{equation}
Away from resonance lines, the Fermi functions can be approximated by step functions. Setting
$f^+_L=f^-_R=1$, $f^-_L=f^+_R=0$ (valid at positive electrochemical potential $eV_\mathrm{b}$), the Liouvillian simplifies to
\begin{equation}
	\mathcal{L}_t
	=
	\frac{\Gamma_0}{3}
	\begin{pmatrix}
		 -1 && 0 && 2 && \frac{Y}{2} && \frac{Y^*}{2}		 	\\
		 0 && -1&& 2 && \frac{Y}{2} && \frac{Y^*}{2}		 	\\
		 1 && 1&& -4 && -Y && -Y^*								\\
		 \frac{Y^*}{2} && \frac{Y^*}{2} && 2(1+Y^*) && -1 && 0	\\
		 \frac{Y}{2} && \frac{Y}{2} && 2(1+Y) && 0 && -1
	\end{pmatrix},
\end{equation}
where $Y=e^{i 2\pi/3}$. 
Similarly, the density matrix becomes
\begin{equation}
	\rho^\infty
	=
	\frac{1}{2}\left(1, 1, 0, e^{-i2\pi/3}, e^{i2\pi/3} \right)^\top,
\end{equation}
which shows full occupation of the $5$--particles groundstates and an empty $6$--particles state. Therefore the current through the system is blocked.
Since the $5_0$ groundstates block the current, one speaks of groundstates blockade \cite{Donarini2009}.
The stability diagram for the current, Fig.~\ref{fig2}(a), shows how the current gets strongly suppressed at the groundstates blockade and only features a small line of finite current at the groundstates resonance, $E_6(V_\mathrm{g},V_\mathrm{b})=E_{5_0}(V_\mathrm{g},V_\mathrm{b})$.

As expected, the Fano factor is Poissonian, $F=1$, in the region of the $6$--particles Coulomb diamond and diverges for $V_\mathrm{b}\to 0$.
In the region of the groundstates blockade it has the super-Poissonian value of $F_{nv}=5/3$, in agreement with the full numerical results shown in Fig.~\ref{fig2}(c).

\section{Current and Fano factor for a minimal model at $\xi=-7.5\vert b \vert$}
\label{sec:A-4/3}
A striking feature at the left side of the stability diagram, a Fano factor of $F=4/3$, cannot be obtained considering a minimal model using only the $5$-- and $6$--particle states. It appears at bias and gate voltages where $f^-_R(E_{5_0}-E_{4_0})$ overcomes $f^+_R(E_6-E_{5_0})$ and the transition to the triplet of ground states with $4$ particles becomes the new bottleneck of transport. At $\xi=-7.5\vert b \vert$, this happens at $eV_\mathrm{b}=(E_6 - E_{4_0})= V =2\vert b \vert$. A minimal model can be written in the basis $\{\vert 6 \rangle\langle 6 \vert, \vert d \rangle\langle d \vert, \vert c \rangle\langle c \vert, \vert 4 \rangle\langle 4 \vert \}$, with the coupled $\vert c \rangle$, decoupled $\vert d \rangle$ and the channel $\vert 4 \rangle$ associated to the triplet $\{4, E_{4_0}; 1,S_z,0\rangle\}$. The Liouvillian and current superoperators are
\begin{widetext}
\begin{align}
	\mathcal{L}
	=&
	\begin{pmatrix}
		-\Gamma^c_{R,65} - \Gamma^d_{R,65}	& 0 						& \Gamma^c_{L,65} 	& 0 \\
		\Gamma^d_{R,65}						& -\Gamma^d_{R,54} f^-_R	& 0					& \Gamma^d_{L,54} + \Gamma^d_{R,54} \\
		\Gamma^c_{R,65}						& 0							& -\Gamma^c_{L,65}	& \Gamma^c_{L,54} + \Gamma^c_{R,54} \\
		0									& \Gamma^d_{R,54} f^-_R		& 0					& -\Gamma^d_{L,54} - \Gamma^d_{R,54} - \Gamma^c_{L,54} - \Gamma^c_{R,54}
	\end{pmatrix}
	,& \\
	\mathcal{J}^+
	=&
	\begin{pmatrix}
		0				& 0 					& 0 & 0 \\
		\Gamma^d_{R,65}	& 0						& 0	& 0 \\
		\Gamma^c_{R,65}	& 0						& 0	& 0 \\
		0				& \Gamma^d_{R,54} f^-_R	& 0	& 0
	\end{pmatrix}
	,\qquad\qquad
	\mathcal{J}^-
	=
	\begin{pmatrix}
		0 & 0 & 0 & 0 				\\
		0 & 0 & 0 & \Gamma^d_{R,54} \\
		0 & 0 & 0 & \Gamma^c_{R,54} \\
		0 & 0 & 0 & 0
	\end{pmatrix}
	,&
\end{align}
\end{widetext}
where $f^-_R$ is the Fermi function between the $5$- and $4$-particle groundstates at the right lead and is responsible for the bottleneck process.
To lowest order in this Fermi function, the current and Fano factor read
\begin{widetext}
\begin{equation}
\begin{split}
	I
	&=
	-e \Gamma^d_{R,54} f^-_R \frac{
		\Gamma^c_{L,54} (\Gamma^c_{R,65}+2 \Gamma^d_{R,65})+\Gamma^c_{R,54} (\Gamma^c_{R,65}+\Gamma^d_{R,65})+\Gamma^d_{L,54} \Gamma^d_{R,65}
		}{
		\Gamma^d_{R,65} (\Gamma^c_{L,54}+\Gamma^c_{R,54}+\Gamma^d_{L,54}+\Gamma^d_{R,54})
	}
	, \\
	F_{nv}
	&=
	\frac{
		\Gamma^c_{L,54} \left(2 (\Gamma^c_{R,65})^2+5 \Gamma^c_{R,65} \Gamma^d_{R,65}+4 (\Gamma^d_{R,65})^2\right)+\Gamma^c_{R,54} \left(2 (\Gamma^c_{R,65})^2+3 \Gamma^c_{R,65} \Gamma^d_{R,65}+(\Gamma^d_{R,65})^2\right)+\Gamma^d_{L,54} (\Gamma^d_{R,65})^2
		}{
		\Gamma^d_{R,65} (\Gamma^c_{L,54} (\Gamma^c_{R,65}+2 \Gamma^d_{R,65})+\Gamma^c_{R,54} (\Gamma^c_{R,65}+\Gamma^d_{R,65})+\Gamma^d_{L,54} \Gamma^d_{R,65})
	}.
	\label{eq:IF_456}
\end{split}
\end{equation}
\end{widetext}
The rate matrix for the $5_0 \leftrightarrow 4_0$ transitions in the angular momentum basis reads $(\mathcal{R}_\alpha)_{\ell\ell^\prime}^{5_0\leftrightarrow 4_0} = \frac{1}{2}\sum_{\sigma,\tau} \langle 5,E_{5_0};\frac{1}{2} , \sigma, \ell|d_{\alpha \tau}^\dagger\mathcal{P}_{4,E_{4_0}} d_{\alpha \tau }|5,E_{5_0}; \frac{1}{2}, \sigma, \ell^\prime\rangle$, where $\mathcal{P}_{NE}=\sum_{S_z,L_z}\vert N,E;S,S_z,L_z \rangle\langle N,E;S,S_z,L_z \vert$ is the projector on the $N$-particle level with energy $E$ and spin $S$ and $\ell,\ell^\prime=\pm 1$. Under the bias and gate voltage conditions considered here, the system still remains in the interference groundstate blocking associated to the $5_0 \leftrightarrow 6$ transition. 
The corresponding coupled-decoupled basis introduced shortly above Eq.~(\ref{eq:rate_matrices}) is thus the most convenient representation. With the help of the eigenstates listed in Table~\ref{tab2} and the Wigner-Eckart theorem one calculates, $\Gamma^d_{\alpha,\, 54} = \frac{3}{2}\Gamma R_\alpha^f$ and $\Gamma^c_{\alpha,\, 54} = \frac{3}{2}\Gamma R_\alpha^s$ with $R_\alpha^{f/s}$ as given below Eq.~(\ref{eq:rate_matrices}). Finally, by substitution into Eq.~(\ref{eq:IF_456}), one obtains $I=-e\Gamma f^-_R/4$ and $F=4/3$.

%\bibliography{bib}
%

\end{document}